\documentclass{aa}  
\usepackage{graphics}
\usepackage{txfonts}
\usepackage[colorlinks=true,citecolor=blue]{hyperref}
\usepackage{caption}

\newcommand{\heii}{He~\textsc{ii}~}
\newcommand{\lya}{Ly$\alpha$~}
\newcommand{\civ}{C~\textsc{iv}~}
\newcommand{\ciii}{C~\textsc{iii}]~}
\newcommand{\oiii}{O~\textsc{iii}]~}

\newcommand{\flux}{erg~s$^{-1}$~cm$^{-2}$~}

\usepackage{color}
\usepackage[normalem]{ulem}

\begin{document}

   \title{The properties of \heii $\lambda1640$ emitters at $z\sim2.5-5$ from the VANDELS survey}
   
   \titlerunning{\heii emitters in VANDELS}


   \author{A. Saxena\inst{1}
          \and
          L. Pentericci\inst{1}
          \and
          M. Mirabelli\inst{1,2}
          \and
          D. Schaerer\inst{3}
          \and
          R. Schneider\inst{1,2,4}
          \and
          F. Cullen\inst{5}
          \and
          R. Amorin\inst{6,7}
          \and
          M. Bolzonella\inst{8}
          \and
          A. Bongiorno\inst{1}
          \and
          A. C. Carnall\inst{5}
          \and
          M. Castellano\inst{1}
          \and
          O. Cucciati\inst{8}
          \and
          A. Fontana\inst{1}
          \and
          J. P. U. Fynbo\inst{9}
          \and
          B. Garilli\inst{10}
          \and
          A. Gargiulo\inst{10}
          \and
          L. Guaita\inst{11}
          \and
          N. P. Hathi\inst{12}
          \and
          T. A. Hutchison\inst{13,14}
          \and
          A. M. Koekemoer\inst{12}
          \and
          F. Marchi\inst{1}
          \and
          D. J. McLeod\inst{5}
          \and
          R. J. McLure\inst{5}
          \and
          C. Papovich\inst{13,14}
          \and
          L. Pozzetti\inst{8}
          \and
          M. Talia\inst{8}
          \and
          G. Zamorani\inst{8}
          }

   \institute{INAF -- Osservatorio Astronomico di Roma, via Frascati 33, I-00078 Monteporzio Catone, Italy
   \and
        Dipartimento di Fisica, Sapienza Universit\`{a} di Roma, Piazzale Aldo Moro 5, 00185 Roma, Italy 
        \and
        Geneva Observatory, University of Geneva, ch. des Maillettes 51, 1290 Versoix, Switzerland
        \and
        INFN, Sezione di Roma I, Piazzale Aldo Moro 2, 00185 Roma, Italy
        \and
        SUPA (Scottish Universities Physics Alliance), Institute for Astronomy, University of Edinburgh, Royal Observatory, EH9 3HJ Edinburgh, UK
        \and
        Instituto de Investigaci\'{o}n Multidisciplinar en Ciencia y Tecnolog\'{i}a, Universidad de La Serena, Ra\'{u}l Bitr\'{a}n 1305, La Serena, Chile
        \and
        Departamento de F\'{i}sica y Astronom\'{i}a, Universidad de La Serena, Av. Juan Cisternas 1200 Norte, La Serena, Chile
        \and
        INAF -- OAS Bologna, Via P. Gobetti 93/3, I-40129, Bologna, Italy
        \and
        Cosmic DAWN Center, Niels Bohr Institute, University of Copenhagen, Juliane Maries Vej 30, 2100 Copenhagen \O, Denmark
        \and
        INAF -- IASF Milano, via Bassini 15, I-20133 Milano, Italy
        \and
        N\'{u}cleo de Astronom\'{i}a, Facultad de Ingenier\'{i}a, Universidad Diego Portales, Av. Ej\'{e}rcito 441, Santiago, Chile
        \and
        Space Telescope Science Institute, 3700 San Martin Drive, Baltimore, MD 21218, USA
        \and
        Department of Physics and Astronomy, Texas A\&M University, College Station, TX, 77843-4242 USA
        \and
        George P. and Cynthia Woods Mitchell Institute for Fundamental Physics and Astronomy, Texas A\&M University, College Station, TX, 77843-4242 USA
    \\
   \email{aayush.saxena@inaf.it}
             }

   \date{Received 22 November 2019 / Accepted 05 March 2020}

 
  \abstract
   {}
   {Strong \heii emission is produced by low-metallicity stellar populations. Here, we aim to identify and study a sample of \heii $\lambda 1640$-emitting galaxies at redshifts of $z \sim 2.5-5$ in the deep VANDELS spectroscopic survey.}
   {We identified a total of 33 \emph{Bright} \heii emitters (S/N > 2.5) and 17 \emph{Faint} emitters ($\textrm{S/N} < 2.5$) in the VANDELS survey and used the available deep multi-wavelength data to study their physical properties. After identifying seven potential AGNs in our sample and discarding them from further analysis, we divided the sample of \emph{Bright} emitters into 20 \emph{Narrow} (FWHM < 1000 km s$^{-1}$) and 6 \emph{Broad} (FWHM > 1000 km s$^{-1}$) \heii emitters. We created stacks of \emph{Faint}, \emph{Narrow,} and \emph{Broad} emitters and measured other rest-frame UV lines such as \oiii and \ciii in both individual galaxies and stacks. We then compared the UV line ratios with the output of stellar population-synthesis models to study the ionising properties of \heii emitters.}
   {We do not see a significant difference between the stellar masses, star-formation rates, and rest-frame UV magnitudes of galaxies with \heii and no \heii emission. The stellar population models reproduce the observed UV line ratios from metals in a consistent manner, however they under-predict the total number of \heii ionising photons, confirming earlier studies and suggesting that additional mechanisms capable of producing He \textsc{ii} are needed, such as X-ray binaries or stripped stars. The models favour subsolar metallicities ($\sim0.1Z_\odot$) and young stellar ages ($10^6 - 10^7$ years) for the \heii emitters. However, the metallicity measured for \heii\  emitters is comparable to that of non-\heii emitters at similar redshifts. We argue that galaxies with \heii emission may have undergone a recent star-formation event, or may be powered by additional sources of \heii ionisation.}
   {}


   \maketitle
%

\section{Introduction}
Understanding the nature of the key drivers of reionisation, a process through which the intergalactic medium (IGM) in the Universe made a phase transition from neutral to completely ionised (by $z\sim6$), is one of the most exciting challenges in cosmology today. The general consensus now is that reionisation was predominantly driven by low-mass star-forming galaxies at $z>6$ \citep{rob10, rob15, bou15}. The key requirement from such galaxies is the production and escape of a sufficiently large number of photons with energies $E>13.6$ eV that are capable of ionising all the neutral hydrogen in the Universe. Galaxies with low metallicities and a high escape fraction of these ionising photons ($\ge10$\%) should in principle be able to complete the reionisation process by $z\sim6$ \citep[see][for example]{sta16}. Recently, \citet{fin19} explored scenarios in which galaxies with low ionising photon escape fractions ($<5\%$) could potentially ionise the IGM.

There is growing observational evidence that galaxies in the early Universe ($z>2$) have lower metallicities, both in the gas phase and in stars, compared to the local galaxy population that we see today \citep{erb10, hen13, ste14, cul14, cul19, amo17, san18}. Lower metallicities naturally lead to the production of harder UV ionisation fields \citep{ste14, ste16}. From a theoretical point of view, the first galaxies and proto-galaxies formed in the Universe must essentially consist of metal-free (or Pop-III) stars that are capable of producing a large number of UV ionising photons that would ultimately be responsible for the very early stages of reionisation \citep{tum00, bar01, sok04, rob10, bro11, wis12}.

One important consequence of the presence of metal-free gas in the early Universe is that stars formed out of it can have very high masses and temperatures \citep[e.g.][]{bro04, bro11}, leading to a top-heavy initial mass function (IMF). With such an IMF, a large number of very high-energy photons can be produced. Such photons should  in principle be capable of ionising UV emission lines requiring very high ionisation energies, such as \heii $\lambda1640$ (where the He$^{+}$ ionisation potential is $>54.4$ eV, $\lambda<228$ \AA) and give rise to strong line fluxes \citep{sch02}. 

The link between strong \heii emission and low metallicities has been shown to exist both using theory and observations at a broad range of redshifts. \citet{sch03} showed that strong \heii accompanied by strong \lya are expected in the spectra of galaxies that are transitioning from Pop-III dominated to more `normal' galaxies that have already been observed. In the local Universe, it has been shown that extremely metal-poor dwarf galaxies often have strong \heii $\lambda4686$ emission in their spectra \citep{gus00, izo04, shi12, keh15, keh18, sen17, sen19, ber19}. In individually detected lensed galaxies with low metallicities at high redshifts, detection of \heii $\lambda1640$ has also been reported \citep{pat16, ber18}. 

Deep spectroscopic surveys of galaxies at $z>2$ have also led to a number of detections of the \heii $\lambda1640$ line. \citet{cas13} argued the case of a possible contribution from Pop III-like stars in order to explain the narrow \heii emission lines observed in their sample of star-forming galaxies at $z\sim2-4.6$. For their broad \heii emitters, the most likely explanation for powering the emission was either winds driven from Wolf-Rayet (WR) stars \citep[see][for example]{sch96, shi12} or a contribution from the accretion disc of active galactic nuclei (AGNs), as is seen in broad \heii-emitting galaxies at low redshifts. \citet{nan19} showed that to explain the observed line ratios and line fluxes for \heii emitters seen in star-forming galaxies at $z\sim2-4$, UV emission line diagnostic tests point towards ionisation by stellar populations with subsolar metallicities.

In addition to low-metallicity stars, several other physical mechanisms have been proposed in the literature that could lead to strong \heii emission in the spectra of galaxies. For example, inclusion of stellar rotation, which captures the effects of rotational mixing leading to higher effective temperatures and moderate mass-loss rates may be able to explain the observed \heii emission in low-metallicity galaxies \citep{szc15}. The models including interacting binary stars, such as Binary Population and Spectral Synthesis\footnote{\url{https://bpass.auckland.ac.nz}} (BPASS; \citealt{eld17, sta18, xia18}), improve overall spectral fits and observed emission line ratios in star-forming galaxies at high redshifts, predominantly due to higher stellar effective temperatures producing harder ionising spectra for an extended period of time. These effects help in better reproducing features such as the narrow nebular \heii emission line in comparison to the single star models \citep[see][for example]{ste16}. However, binary star models fall short of reproducing the observed \heii equivalent widths in high-redshift galaxies, even after considering different initial mass functions in the modelling \citep{sta19, nan19}. Careful modelling of mass transfer in binaries that may lead to `stellar stripping' and rejuvenation of old stars may also provide the extra \heii-ionising photons needed to explain the observed emission lines \citep{got18, got19}.

An additional contribution from X-ray binaries (XRBs) has been suggested as one of the mechanisms that may be capable of explaining the \heii line fluxes observed in low-metallicity galaxies, both in the local Universe and at high redshifts. It has been shown that the X-ray luminosities of star-forming galaxies increase with decreasing metallicities, primarily because of an enhanced contribution from high-mass XRBs at lower metallicities \citep{fra13a, fra13b, dou15, bro16}. \citet{sch19} showed that including the contribution from XRBs in addition to single or binary stellar populations can explain the \heii emission line strength and its dependence on metallicity in local low-metallicity galaxies. However, \citet{sen19b} argue that high-mass XRB populations may not be sufficient to account for the observed \heii line strengths in nearby metal-poor galaxies, and revised stellar wind models or inclusion of softer X-ray sources may be needed. At higher redshifts, \citet{for19} showed that X-ray observations of stacks of star-forming galaxies are consistent with the metallicity dependence of the XRB populations at low redshifts. Given the overall decrease in metallicity of galaxies reported at high redshifts, it is expected that the contribution from XRBs plays an increasingly important role in the spectra of high-redshift galaxies, and the \heii line offers a probe for this.

Developing a better understanding of the nature of galaxies capable of producing very high-energy ionising photons is vital, and in this paper we present a new sample of \heii $\lambda1640$-emitting galaxies identified in the VANDELS survey. In Section \ref{sec:sample} we present details of our sample selection, briefly describing the VANDELS survey, and present our \heii identification procedure. We also describe our line measurement and stacking techniques. In Section \ref{sec:physicalprops} we discuss key physical properties of \heii-emitting galaxies, comparing them with results from other deep surveys at comparable redshifts in the literature. In Section \ref{sec:models} we explore the observed UV line ratios of both individual \heii emitters as well as stacks of galaxy spectra, comparing them with photo-ionisation models including both single-star and binary-star models. In Section \ref{sec:discussion} we discuss the properties of \heii-emitting galaxies in the context of the overall population of galaxies at similar redshifts and discuss some possible mechanisms that could give rise to \heii emission. Finally, in Section \ref{sec:conclusions} we summarise the key findings of this study.

Throughout this paper, we assume a flat $\Lambda$CDM cosmology with $\Omega_\textrm{m} = 0.3$, $\Omega_\Lambda = 0.7$ and H$_0 = 67.7$ km s$^{-1}$ Mpc$^{-1}$ taken from \citet{planck}. We use the AB magnitude system \citep{oke83} throughout this paper.


\section{Sample selection and \heii emitter identification}
\label{sec:sample}
In this study, we use data from VANDELS,  a deep VIMOS survey of the CDFS and UDS fields in CANDELS \citep{gro11, koe11}, which itself is  a recently completed ESO public spectroscopic survey carried out using the \textit{Very Large Telescope (VLT)}. Details about the survey description and initial target selection can be found in \citet{mcl18} and details about the data reduction and redshift determination can be found in \citet{pen18}. The targets for VANDELS are selected from two well-studied extragalactic fields, the UKIDSS Ultra Deep Survey (UDS) centred around RA = 02:17:38, Dec = $-$05:11:55, and the Chandra Deep Field South (CDFS) centred around RA = 03:32:30, Dec = $-$27:48:28. In the following we briefly describe the details of the survey that are relevant to this work.

The VANDELS survey contains spectra of approximately $2100$ galaxies in the redshift range $1.0<z<7.0$, with on-source integration times ranging from 20 to 80 hours, where over $70\%$ of the targets have at least 40 hours of integration time. The galaxies targeted for spectroscopy include the star-forming galaxy population at $z>2.4$ and massive, passive galaxies in the redshift range $1<z<2.5$. The redshift ranges are primarily dictated by the sensitivity of the VIMOS grism used for VANDELS observations \citep{mcl18}, which has a wavelength coverage in the range $4800-10000$ \AA. The resolution provided by these observations is $R\sim600$. The final spectra have high signal-to-noise ratios (S/Ns), enabling detailed absorption and emission line studies, the study of ionising sources using emission line ratios, the derivation of accurate metallicities, and better constraints on physical parameters such as stellar masses and star-formation rates. 

The spectra were reduced using the \textsc{Easylife} data reduction pipeline \citep{gar12} and data products delivered by VANDELS consist of the extracted 1D spectra, the 2D re-sampled and sky-subtracted spectra, and catalogues with essential galaxy parameters, including spectroscopic redshifts. The reliability of redshifts in the VANDELS database is recorded by four individual and independent measurers from the VANDELS team. Based on how confident the individual measurers were about the accuracy of the redshift, the following flags were assigned (which are important for this work): 0 = no redshift could be assigned; 1 = 50\% probability of being correct; 2 = 70-80\% probability of being correct; 3 = 95-100\% probability of being correct; 4 = 100\% probability of being correct; and 9 = spectrum shows a single emission line. The typical accuracy of spectroscopic redshift measurements is $\sim150$ km~s$^{-1}$. More details about the VANDELS data products as well as redshift determination and flag assignment can be found in \citet{pen18}.

For this work, we initially selected galaxies from the VANDELS database in the following way. We selected spectra of only those star-forming galaxies that have a redshift reliability flag of either 3 or 4, which guarantees that the redshift measured by the VANDELS team has a more than $95$\% probability of being correct. This also ensures that there must be multiple emission or absorption features visible and that the spectrum has a high S/N. From these, we select sources in the redshift range $2.2<z<4.8$, which ensures that the \heii $\lambda 1640$ emission line, if present in the galaxy spectra, lies in regions of high throughput. From the UDS field, 455 sources satisfied the selection criteria and from CDFS, 494 sources satisfied the selection criteria. We refer to this total sample of 949 galaxies as the `parent' sample.

\subsection{Line identification}
The spectra of galaxies in the parent sample were then inspected for the presence of any \heii emission. Since we have only selected sources with highly reliable spectroscopic redshifts, we visually inspected the wavelength region where the \heii line is to be expected using the \textsc{pandora} software\footnote{\url{http://cosmos.iasf-milano.inaf.it/pandora/}}, which is a suite of \textsc{python} modules designed for visualising and analysing spectra, amongst other useful operations. In cases where the \heii line coincided with a strong skyline feature, we discarded the spectrum as the emission line measurement would not be reliable. Skyline contamination is particularly dominant at \heii redshifts $4.0 < z < 4.5$, resulting in lower numbers of shortlisted sources in this range. For seemingly fainter \heii emission, the 2D spectra were also inspected to look for signatures of emission and to distinguish between real signal and hot pixels or weak sky residuals. The initial identification process based on 1D and 2D spectra was carried out independently by two individuals, and only those sources that are identified as \heii emitters by both individuals were retained for further analysis. A final visual examination of the shortlisted \heii emitters was then independently carried out by a third individual from our team.

The sources identified to have \heii emission were then moved on to the next stage of the analysis, where the \heii line (and any other strong line present in the spectrum) was measured. The line fitting is performed using a single Gaussian at the position of the \heii emission line in the observed frame using \emph{MUSE Python Data Analysis Framework} or \textsc{mpdaf}\footnote{\url{https://mpdaf.readthedocs.io/en/latest/index.html}}, which provides tools to analyse spectra, images, and data cubes. A S/N of the integrated line flux is obtained, and the redshifts of sources with \heii (or C \textsc{iii}]) emission are updated, as \heii (or C \textsc{iii}]) is a more accurate tracer of the systemic redshift \citep[see also][]{nan19}. When both lines are present in the spectrum, we use the \heii redshift. However, we generally find a reasonable agreement between the redshifts measured using \heii and C \textsc{iii}]. At this stage, the sources are separated into two subsamples based on the inferred S/N of the emission line (similar to \citealt{nan19}). The subsample of \emph{Bright} \heii emitters  contains galaxies with a S/N of the \heii line $>2.5$, and the subsample of \emph{Faint} \heii emitters  contains galaxies with \heii S/N $<2.5$. In the \emph{Faint} sample, the lowest measured S/N of the \heii emission line is $1.5$.

In the CDFS field, out of a total of 494 sources in the parent sample, 26 were shortlisted as \heii emitters after the visual check. Of these, 19 are \emph{Bright} emitters and 7 are \emph{Faint} emitters. In the UDS field, the total number of sources in the parent sample was 455, with 24 \heii emitters, out of which 14 are \emph{Bright} and 10 are \emph{Faint}. Across both fields, this leaves us with a total of 33 \emph{Bright} emitters and 17 \emph{Faint} emitters.

\subsection{Line measurements of Bright sources}
The individual spectra of sources in the \emph{Bright} \heii emitters sample were re-analysed using a more careful line-fitting approach. The observed spectrum of each source is first converted to rest frame using the redshift determined by the peak of the \heii line which was measured earlier. The rest-frame continuum is then estimated by fitting a high-degree polynomial to the sigma-clipped spectrum around the \heii emission line. The continuum is subtracted and a Gaussian is fit to the emission line, which yields an integrated flux and full-width at half maximum (FWHM) with associated errors on the fit. The integrated flux is only calculated over the wavelength range where the flux is above the continuum value. This helps to minimise the impact of absorption features close to the emission line on the total flux measured. The equivalent width (EW) of the emission line is also calculated using the measured continuum.
\begin{figure}
        \centering
        \includegraphics[scale=0.45]{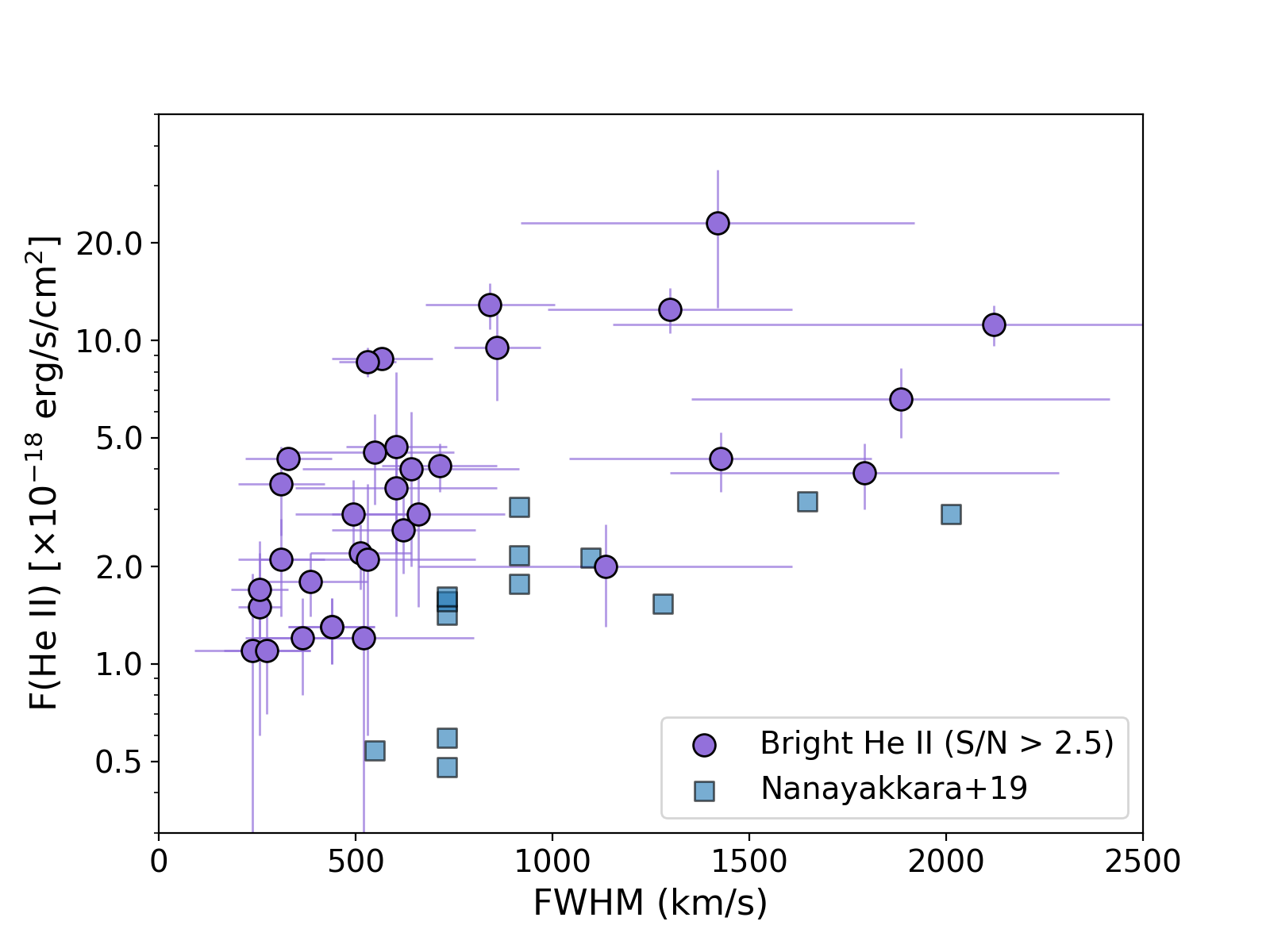}
        \caption{Distribution of the measured \heii line flux and the rest frame FWHM for \emph{Bright} \heii emitters (S/N > 2.5) identified in both CDFS and UDS fields. Shown for comparison are line measurements of \heii emitters from \citet{nan19} at comparable redshifts to our sample.}
        \label{fig:flux_fwhm}
\end{figure}

The \heii line flux in our sample ranges from $1.1\times10^{-18}$ to $3.1\times10^{-17}$ \flux. The FWHM (rest-frame) ranges from 240 to 2120 km s$^{-1}$. The range of EW (rest-frame) measured across galaxies in both fields ranges from 0.9 $\AA$ to 21.4 $\AA$. The measured \heii line properties for the \emph{Bright} sources are shown in Table \ref{tab:strong_sources}. In Figure \ref{fig:flux_fwhm} we show the distribution of the measured \heii fluxes and FWHM (rest frame) for \emph{Bright} sources, and show measurements from the \citet{nan19} sample, which offers the best sample to compare our \heii emitters owing to its very similar redshift range. In Figure \ref{fig:CDFS_spectra} we show the \emph{Bright} \heii emitters identified in both the CDFS and UDS fields. In the following, we identify and remove possible AGNs.

\begin{table*}
\centering
\caption{Measured \heii line properties (rest frame) and derived physical properties of \emph{Bright} sources in CDFS and UDS fields. Based on the width of the \heii line, the sources are grouped as \emph{Narrow} and \emph{Broad}, with the distinction at FWHM(\heii) = 1000 km s$^{-1}$.}
\begin{tabular}{l c c c c c c c c c c r}
\hline
   {ID} & {RA} & {Dec} & {Flux} & {FWHM} & {EW} & {$z_{\textrm{sys}}$} & {$\log_{10}(\textrm{M}_\star$)} & {$\log_{10}$(SFR)}  \\
     & & & \tiny{($\times10^{-18}$ erg~s$^{-1}$~cm$^{-2}$)}  & \tiny{(km s$^{-1}$)}  & \tiny{($\AA$)} &  & \tiny{($M_\odot$)} & \tiny{($M_\odot$ yr$^{-1}$)} \\
\hline \hline
  \emph{Narrow} \\
  \hline
  
  CDFS009705 & 03:32:20.9 & $-$27:49:16.1 & 2.2 $\pm$ 0.5 & 510 $\pm$ 130 & 1.3 $\pm$ 0.4 & 2.484 & 8.9 & 1.2 \\
  CDFS015347 & 03:32:13.2 & $-$27:46:42.6 & 1.3 $\pm$ 0.7 & 240 $\pm$ 145 & 1.9 $\pm$ 1.3 & 3.514 & 9.4 & 1.1 \\
  CDFS019872 & 03:32:47.9 & $-$27:44:29.4 & 1.1 $\pm$ 0.4 & 280 $\pm$ 110 & 1.8 $\pm$ 0.9 & 3.453 & 9.3 & 1.0 \\
  CDFS023170 & 03:32:37.8 & $-$27:42:32.5 & 2.8 $\pm$ 1.5 & 455 $\pm$ 110 & 1.1 $\pm$ 0.6 & 2.977 & 9.7 & 2.0 \\
  CDFS023215 & 03:32:40.7 & $-$27:42:34.7 & 4.3 $\pm$ 0.2 & 330 $\pm$ 110 & 4.1 $\pm$ 2.6 & 3.471 & 9.6 & 1.3 \\
  CDFS023527 & 03:32:18.8 & $-$27:42:48.1 & 1.5 $\pm$ 0.9 & 260 $\pm$ 55 & 0.9 $\pm$ 0.4 & 3.108 & 9.4 & 1.7 \\
  CDFS113062 & 03:32:02.6 & $-$27:52:23.7 & 4.0 $\pm$ 2.0 & 640 $\pm$ 275 & 1.4 $\pm$ 0.7 & 2.695 & 9.4 & 1.7 \\
  CDFS122687 & 03:33:00.1 & $-$27:47:32.6 & 2.9 $\pm$ 0.8 & 495 $\pm$ 150 & 1.0 $\pm$ 0.4 & 2.643 & 9.7 & 2.0 \\
  CDFS126819 & 03:31:55.7 & $-$27:45:33.1 & 9.5 $\pm$ 3.0 & 860 $\pm$ 110 & 8.0 $\pm$ 2.0 & 2.818 & 9.7 & 2.0 \\
  CDFS202040 & 03:32:02.5 & $-$27:58:15.7 & 2.6 $\pm$ 0.7 & 620 $\pm$ 180 & 3.7 $\pm$ 1.9 & 3.474 & 9.1 & 1.1 \\
  CDFS229681 & 03:31:59.4 & $-$27:45:46.5 & 1.2 $\pm$ 0.9 & 520 $\pm$ 280 & 1.9 $\pm$ 1.5 & 3.331 & 8.8 & 1.1 \\  
  UDS004017 & 02:17:37.0 & $-$05:15:15.4 & 1.7 $\pm$ 0.5 & 260 $\pm$ 75 & 1.4 $\pm$ 0.6 & 2.389 & 10.0 & 1.6 \\
  UDS013586 & 02:17:52.5 & $-$05:12:04.8 & 3.1 $\pm$ 1.9 & 335 $\pm$ 100 & 0.6 $\pm$ 0.4 & 2.581 & 9.9 & 2.2 \\
  UDS019505 & 02:17:45.9 & $-$05:10:09.1 & 2.1 $\pm$ 1.5 & 530 $\pm$ 275 & 0.7 $\pm$ 0.5 & 2.865 & 9.6 & 1.4 \\
  UDS020089 & 02:17:22.7 & $-$05:09:54.1 & 2.1 $\pm$ 0.7 & 310 $\pm$ 110 & 2.6 $\pm$ 1.4 & 3.218 & 9.4 & 0.9 \\
  UDS021062 & 02:16:58.1 & $-$05:09:35.4 & 1.2 $\pm$ 0.4 & 365 $\pm$ 145 & 1.1 $\pm$ 0.5 & 2.537 & 9.6 & 1.8 \\
  UDS196554 & 02:17:40.8 & $-$05:05:56.3 & 3.5 $\pm$ 1.3 & 600 $\pm$ 255 & 2.4 $\pm$ 1.3 & 2.618 & 10.0 & 2.3 \\
  UDS200677 & 02:18:14.1 & $-$05:04:49.2 & 4.7 $\pm$ 1.0 & 600 $\pm$ 130 & 2.4 $\pm$ 0.7 & 3.587 & 9.8 & 2.1 \\
  UDS281893 & 02:17:11.3 & $-$05:22:17.6 & 2.9 $\pm$ 1.4 & 660 $\pm$ 220 & 1.1 $\pm$ 0.6 & 2.697 & 9.1 & 1.4 \\
  UDS292392 & 02:18:18.4 & $-$05:20:41.4 & 3.6 $\pm$ 1.1 & 310 $\pm$ 110 & 5.5 $\pm$ 4.0 & 4.555 & 8.9 & 1.2 \\
  \hline
  
  \emph{Broad} \\
  \hline
  CDFS021470 & 03:32:21.9 & $-$27:43:38.8 & 3.9 $\pm$ 0.9 & 1800 $\pm$ 500 & 2.6 $\pm$ 0.8 & 2.574 & 9.1 & 1.3 \\
  CDFS102149 & 03:31:59.7 & $-$27:58:02.6 & 2.0 $\pm$ 0.7 & 1130 $\pm$ 480 & 2.0 $\pm$ 0.9 & 2.614 & 9.9 & 2.0 \\
  CDFS129134 & 03:31:59.0 & $-$27:44:23.5 & 11.2 $\pm$ 1.6 & 2120 $\pm$ 970 & 3.8 $\pm$ 0.9 & 3.202 & 9.6 & 1.8 \\
  CDFS141081 & 03:33:10.1 & $-$27:40:48.2 & 4.3 $\pm$ 0.9 & 1425 $\pm$ 385 & 3.5 $\pm$ 1.2 & 2.385 & 9.9 & 1.4 \\
  CDFS231194 & 03:33:02.6 & $-$27:45:01.8 & 12.5 $\pm$ 2.0 & 1300 $\pm$ 530 & 2.2 $\pm$ 0.6 & 3.077 & 9.5 & 1.8 \\
  UDS137388 & 02:17:12.2 & $-$05:22:31.5 & 23.1 $\pm$ 10.5 & 1420 $\pm$ 500 & 4.6 $\pm$ 3.5 & 2.598 & 9.6 & 1.9 \\
  \hline
  
  \emph{AGN} \\
  \hline
  
  CDFS006327$^*$ & 03:32:42.8 & $-$27:51:02.6 & 1.3 $\pm$ 0.3 & 440 $\pm$ 100 & 5.2 $\pm$ 2.8 & 3.493 & 8.8 & 1.1 \\
  CDFS028933$^*$ & 03:32:07.1 & $-$27:50:55.5 & 1.3 $\pm$ 0.3 & 440 $\pm$ 110 & 4.7 $\pm$ 2.1 & 3.787 & 8.8 & 1.0 \\
  CDFS208098$^*$ & 03:32:52.3 & $-$27:55:26.3 & 6.6 $\pm$ 1.6 & 1885 $\pm$ 530 & 3.5 $\pm$ 1.4 & 2.478 & 9.0 & 1.3 \\
  UDS020721$^{\textrm{x}}$ & 02:17:38.1 & $-$05:09:47.1 & 12.9 $\pm$ 2.1 & 840 $\pm$ 165 & 7.5 $\pm$ 2.8 & 2.520 & 10.7 & 2.1 \\
  UDS021234$^*$ & 02:17:41.0 & $-$05:09:31.3 & 1.8 $\pm$ 0.4 & 385 $\pm$ 145 & 3.7 $\pm$ 2.2 & 4.600 & 9.7 & 0.7 \\
  UDS025482$^{*\textrm{,x}}$ & 02:17:03.4 & $-$05:08:04.7 & 4.1 $\pm$ 0.7 & 710 $\pm$ 145 & 4.3 $\pm$ 1.8 & 3.523 & 10.3 & 0.9 \\
  UDS145830$^*$ & 02:18:13.6 & $-$05:20:11.4 & 8.6 $\pm$ 0.9 & 530 $\pm$ 75 & 21.4 $\pm$ 10.6 & 3.210 & 10.7 & 1.0 \\ 
  
\hline\end{tabular} \\

\tablefoot{Possible AGNs have been indicated using an asterisk for those sources with strong \civ emission in their spectra, and an `X' for sources with X-ray detections. Sources with ID $>100000$ have ground-based photometry only. The RA and Dec are in J2000 epoch.}
\label{tab:strong_sources}
\end{table*}

\subsection{Possible AGNs in the sample}
\label{sec:agn}
The high ionisation potential needed for \heii emission can often be achieved by AGNs and therefore it is important to identify possible AGNs that may be present in the shortlisted sample of \heii emitters from VANDELS. We identify possible AGNs by searching for strong \civ emission in the spectra, similar to \citet{nan19}. The motivation for doing this is that the ionisation potential of C$^{++}$ is high (49.9 eV) and generally the \civ emission undergoes strong absorption from stellar winds. Therefore, it is not very common for the spectra of normal star-forming galaxies to show strong \civ in emission \citep[see][]{sha03}, and more often than not, \civ emission is found in the spectra of AGNs and/or radio galaxies.

Therefore, we inspected the spectrum of all the shortlisted \heii emitters (both \emph{Bright} and \emph{Faint}) to look for signs of \civ emission. Those galaxies that show \civ emission have been marked as likely AGNs, indicated by an asterisk in Table \ref{tab:strong_sources}. We remove these sources from the analysis that follows in order to explore the properties of \heii emission from star-formation activity alone. In Section \ref{sec:civ_emitters} we explore sources that show both \heii and \civ emission in more detail.

Additionally, we use the deep X-ray data available in the VANDELS field to identify X-ray AGNs in the sample. The CDFS X-ray catalogue, with $\approx 7$ Ms of exposure time \citep{luo17}, is significantly deeper than the UDS catalogue with $\approx 600$ ks of exposure time \citep{koc18}. We cross-match the RA and Dec of \heii-emitting sources with the X-ray catalogues using a radius of 1 arcsecond to minimise misidentifications. We do not find any cross-matches in the CDFS field, but find two matches in the UDS field, one of which was already identified as a potential AGN due to the presence of strong \civ emission in its spectrum. X-ray-detected sources are marked with ($^\textrm{x}$) in Table \ref{tab:strong_sources}.

\subsection{Stacking}
Since the average S/N of emission lines in individual spectra is relatively weak, we extend our analysis to also include stacks of \heii emitters. In this section we describe the different classes of objects that are stacked together to boost S/N in order to aid the analysis that follows and we begin by describing the stacking procedure.

The stacking is performed by first converting each spectrum to rest-frame. Wherever accurate systemic redshifts could not be determined, for example due to faintness of the \heii line in \emph{Faint} \heii emitters, we use the spectroscopic redshift determined by the VANDELS team based on template fitting of multiple emission and absorption features in the spectrum \citep{pen18}. The stacking method adopted in this study is similar to that used by \citet{mar18}. The rest-frame spectra are first normalised using the mean flux density value in the wavelength range $1460-1540$ $\AA$, and assigned a weight based on the errors on flux in this range, where a higher weight is assigned to sources with lower errors. The spectra are then re-sampled to a wavelength grid ranging from 1200 to 2000 $\AA$, with a step size of 0.567 $\AA$ -- the wavelength resolution obtained at a redshift of 3.5, which is roughly the median redshift of VANDELS sources in this study. The wavelengths that are not covered in the rest-frame spectra are masked on a source-by-source basis. We also mask residual sky lines in the spectra. The spectra are then stacked using an error-weighted averaging procedure. The $1\sigma$ errors on the stacked spectra are calculated using bootstrapping -- we randomly sample and stack the same number of galaxies from the parent sample 100 times, and the dispersion on fluxes thus obtained gives the errors.

\subsubsection{\emph{Faint} sources}
For sources classified as \emph{Faint} \heii emitters, we rely on stacking to boost the signal of the \heii and other lines present in their spectra. Stacking serves as an additional check to ensure that the \emph{Faint} sources are indeed bona fide \heii emitters, which we can confirm by comparing with the stacked spectrum of all the sources in the parent sample that were not classified as \heii emitters. Before stacking, we ensure that none of the sources are classified as possible AGNs.

\begin{figure*}
        \centering
        \includegraphics[scale=0.8]{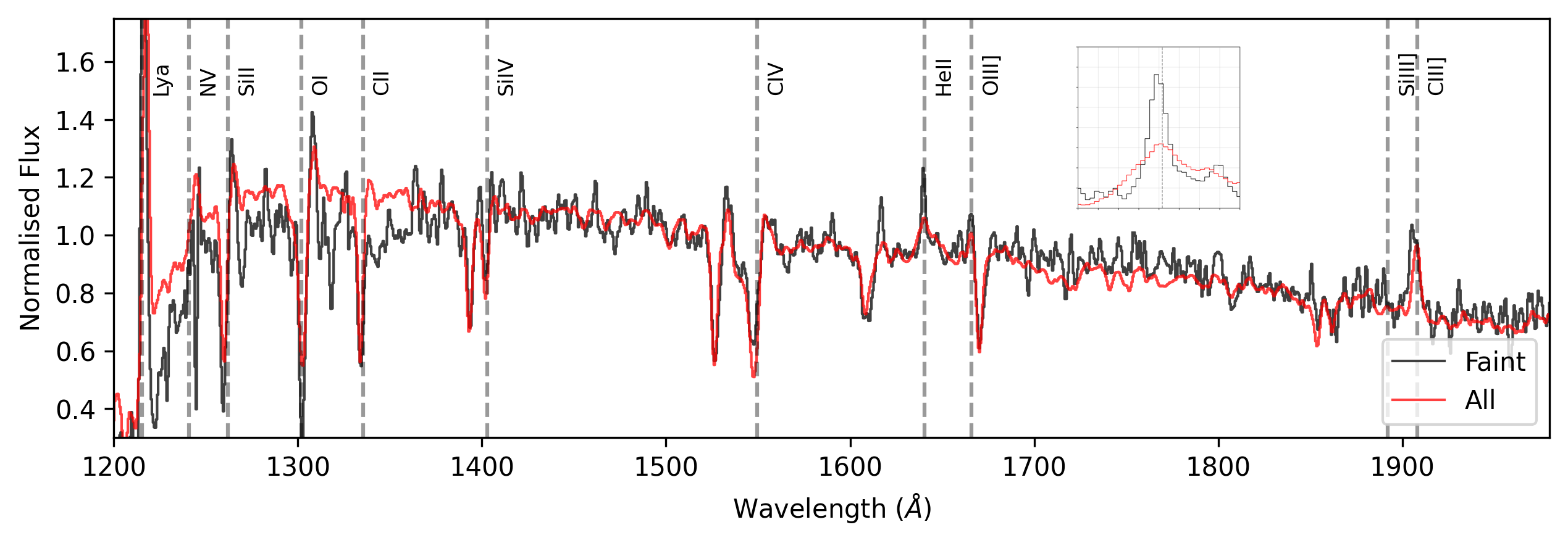}
        \caption{Stacked spectrum of 17 \emph{Faint} \heii emitters from both CDFS and UDS fields (black) and the stack of 899 sources from the parent sample in both fields with no \heii emission (red). Shown in the inset is a zoom-in of the \heii line, clearly demonstrating that there is \heii emission in the stack of the \emph{Faint} emitters. This gives us confidence to include even the \emph{Faint} emitters in this study. The errors on the \emph{Faint} stack are shown in Figure \ref{fig:stacks}.}
        \label{fig:faint_stack}
\end{figure*}

The resulting stack of 17 \emph{Faint} \heii-emitting galaxies in the CDFS and UDS fields (black) and the stack of 899 galaxies not identified as \heii emitters (red) are shown in Figure \ref{fig:faint_stack}, with a zoom-in on the \heii line shown in the inset. The presence of a narrow \heii emission line is clear in the stack of galaxies classified as \emph{Faint} \heii emitters, which gives us confidence to include the \emph{Faint} sources in the analysis that follows. There seems to be weak, possibly broad \heii emission even in the stack of non-\heii-emitting galaxies, which may be resulting from low-level emission present in the spectra of galaxies that we could not identify as \heii emitters with a high degree of confidence. Nevertheless, the higher strength of \heii seen in the stack of \emph{Faint} emitters compared to the non-emitters is clear from Figure \ref{fig:faint_stack}.

\subsubsection{\emph{Broad} and \emph{Narrow} \heii emitters}
To understand the underlying differences between narrow and broad \heii emission lines as well as boost the overall S/N of the spectra of \heii emitters, we produce stacks of sources that fall into each of these classes. Out of the \emph{Bright} emitters, we classify sources with \heii FWHM $< 1000$ km s$^{-1}$ as \emph{Narrow}, and sources with \heii FWHM $> 1000$ km s$^{-1}$ as \emph{Broad}, in line with the analysis of \citet{cas13}. The physical motivation behind such a separation is the underlying mechanism that is likely powering the \heii line. \heii emission from evolved WR stars is expected to be broad, with FWHM $\ge 1000$ km s$^{-1}$ \citep{sch03, bri08}. However, narrow nebular \heii emission is still relatively poorly understood and requires low-metallicity, high-stellar-mass populations. Therefore, to qualitatively separate out the likely WR dominated and low-metallicity populations, we separate sources based on their \heii FWHM. 
\begin{figure}
        \centering
        \includegraphics[scale=0.47]{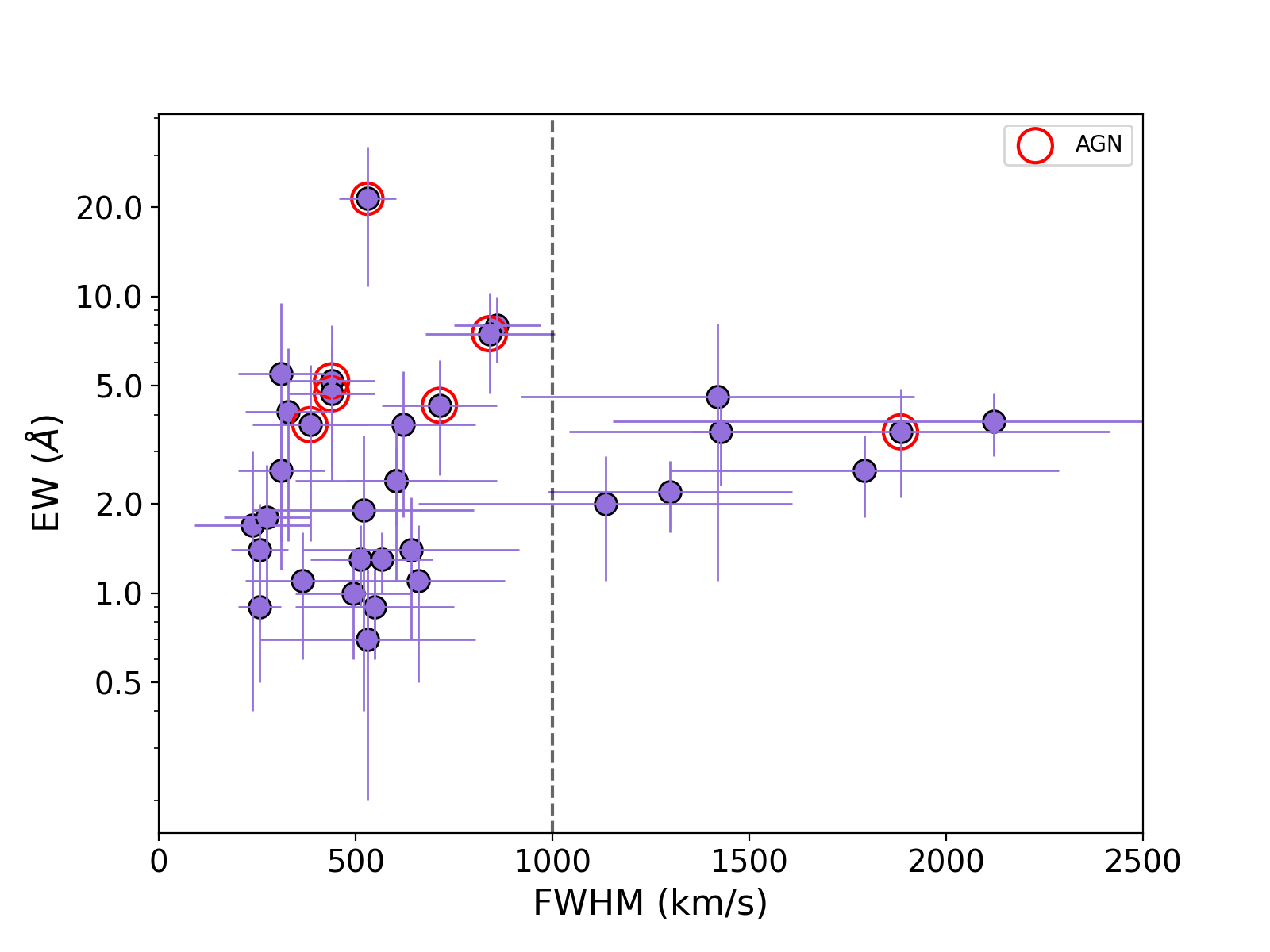}
        \caption{Distribution of \heii EW and FWHM for sources in our sample. Marked by the dashed line is the cut in FWHM = 1000 km s$^{-1}$ introduced to create subsamples for stacking analysis. This separation has been motivated by previously observed (and expected) \heii FWHM from narrow and broad emitters -- the broad \heii can generally be explained by AGN and/or stellar winds, whereas narrow \heii emission is generally attributed to low-metallicity stars \citep[see][]{cas13}. Sources with strong \civ emission or X-ray detection that have been tentatively classified as AGNs are excluded from further analysis, and are marked by red circles.}
        \label{fig:ew_fwhm_bins}
\end{figure} 

\begin{figure*}
        \centering
        \includegraphics[scale=0.8]{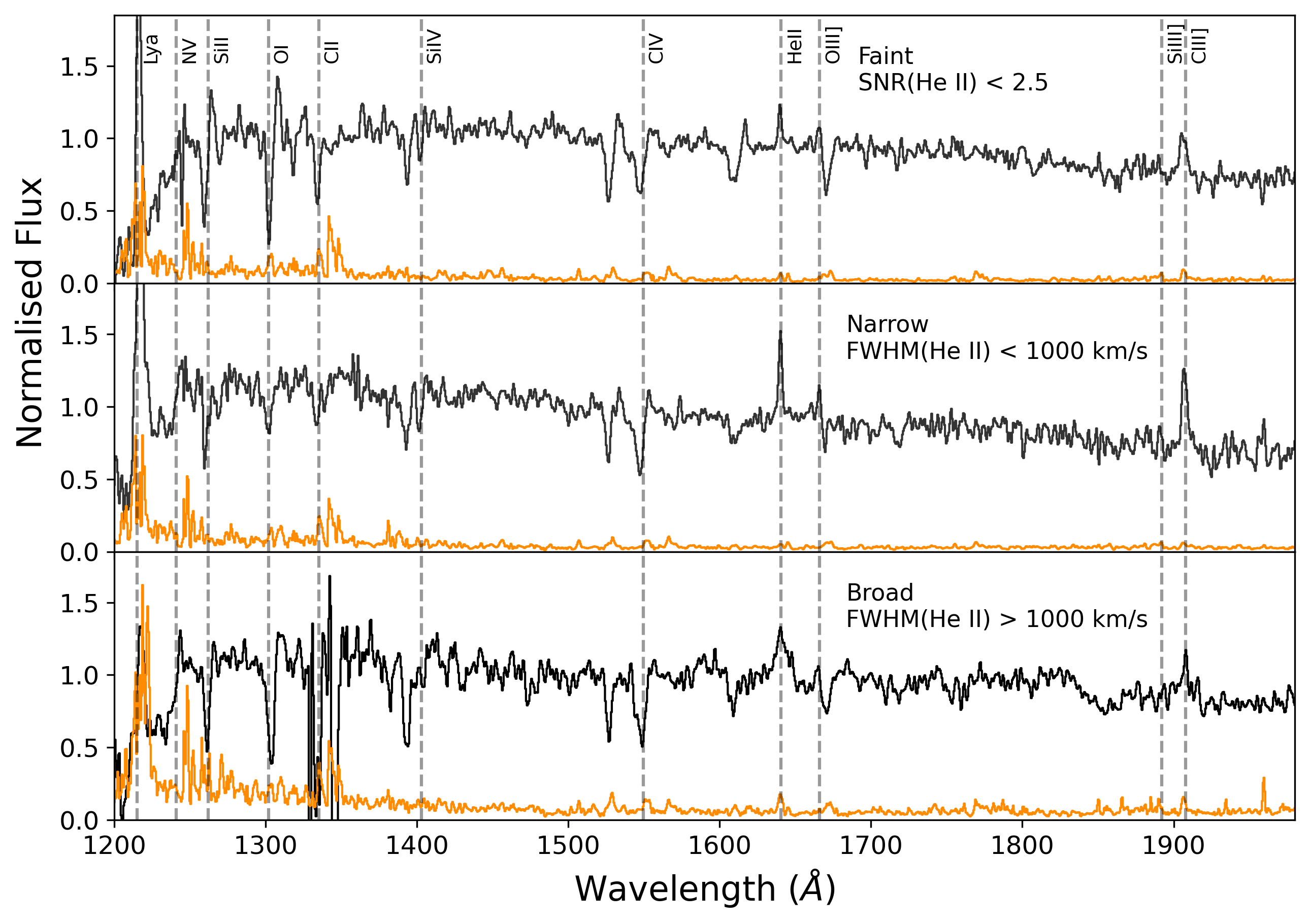}
        \caption{Stacked spectra of sources classified as \emph{Faint} (top), \emph{Narrow} (middle), and \emph{Broad} (bottom) \heii emitters. There are a total of 17 sources making up the \emph{Faint} stack, 20 sources making up the \emph{Narrow}  stack, and 6 sources making up the \emph{Broad} stack. The errors obtained from bootstrapping are shown for each stack.}
        \label{fig:stacks}
\end{figure*}

In Figure \ref{fig:ew_fwhm_bins} we show the distribution of \heii EWs and FWHM of our sample of \emph{Bright} \heii emitters. The dashed line marks the classification of sources as \emph{Narrow} or \emph{Broad}, and red circles mark the sources that show strong \civ emission and/or X-ray detections and are therefore classified as candidate AGNs. After removing seven AGNs from the analysis, we find that 20 \heii emitters are classified as \emph{Narrow} and 6 are classified as \emph{Broad}. The resulting stacked spectra of the two classes of \heii emitters are shown in Figure
\ref{fig:stacks}, together with the stacked spectrum of the \emph{Faint} \heii emitters. Also shown for each stack are errors obtained through bootstrapping. The number of sources that make up the stacked spectrum for each class of \heii emitters is shown in Table \ref{tab:bins}.

\begin{table}
        \centering
        \caption{Number of sources in each subsample used to produce stacked spectra, after removing seven AGNs from the sample.}
        \begin{tabular}{l c r}
        \hline
        Class & Property & No. sources \\
        \hline \hline \\
        \emph{Faint} & S/N(He \textsc{ii}) < 2.5 & 17 \\
        \emph{Narrow} & FWHM(He \textsc{ii}) < 1000 km s$^{-1}$ & 20 \\
        \emph{Broad} & FWHM(He \textsc{ii}) $\ge$ 1000 km s$^{-1}$ & 6 \\
        \hline
        \label{tab:bins}
        
        \end{tabular}
\end{table}

The emission lines in these stacks are then measured by fitting a single Gaussian to the stacked spectrum. Since only line ratios are required for diagnostic plots, the normalisation of the stacked spectra is cancelled out. To capture the errors on the continuum and the subsequent errors on the line flux measurements introduced by stacking, we use the following method. We choose a region in the spectrum that is free of emission close to each line of interest, and we calculate the mean and standard deviation of the continuum close to the line. We then randomly sample 100 continuum values within $\pm 1\sigma$ of the mean continuum. We then fit a Gaussian to the emission line for every continuum value determining the line flux. The median emission line flux is the median of 100 fluxes obtained using varying continuum values, and the standard deviation of the line fluxes gives the error on the flux measurement of the line. 

\section{Physical properties of \heii emitters}
\label{sec:physicalprops}
Thanks to the availability of excellent multi-wavelength data ranging from ultraviolet (UV) to mid-infrared (MIR) from both space-based and ground-based telescopes in the VANDELS fields, it is possible to accurately derive the physical properties of sources with reliable spectroscopic redshifts. Physical parameters such as stellar masses ($M_\star$), dust attenuation ($A_V$), star-formation rates (SFRs), and rest-frame absolute UV magnitudes ($M_{\textrm{UV}}$) were obtained by fitting spectral energy distribution (SED) templates to photometric points from broad-band filters at the spectroscopic redshift of each galaxy. The SED fits were performed using $Z=0.2$ $Z_\odot$ metallicity versions of the standard \citet{bc03} models with redshifts fixed to the VANDELS spectroscopic redshift. The star-formation rates were corrected for dust using the fitted $A_V$ value and adopting the \citet{cal00} dust attenuation law. The rest-frame magnitudes were calculated using a 200 $\AA$ wide top-hat filter centred at 1500 $\AA$. We refer the readers to \citet{mcl18} for full details of the SED fitting techniques, model assumptions, and derived physical parameters.

\subsection{Stellar masses}
\begin{figure*}
        \centering
        \begin{minipage}{0.32\textwidth}
            \centering
                \includegraphics[scale=0.42]{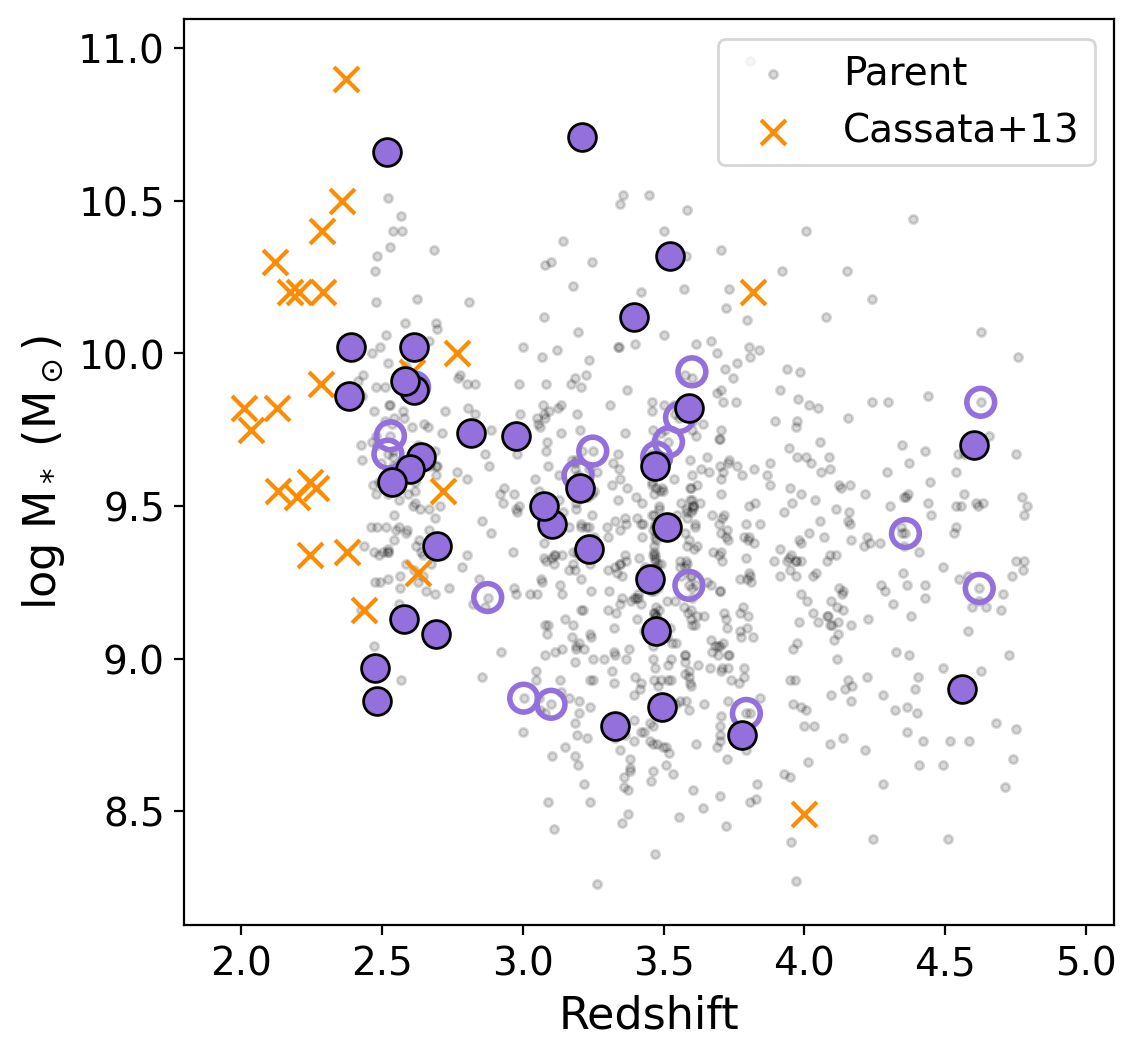}
        \end{minipage}
        \begin{minipage}{0.32\textwidth}
            \centering
                \includegraphics[scale=0.42]{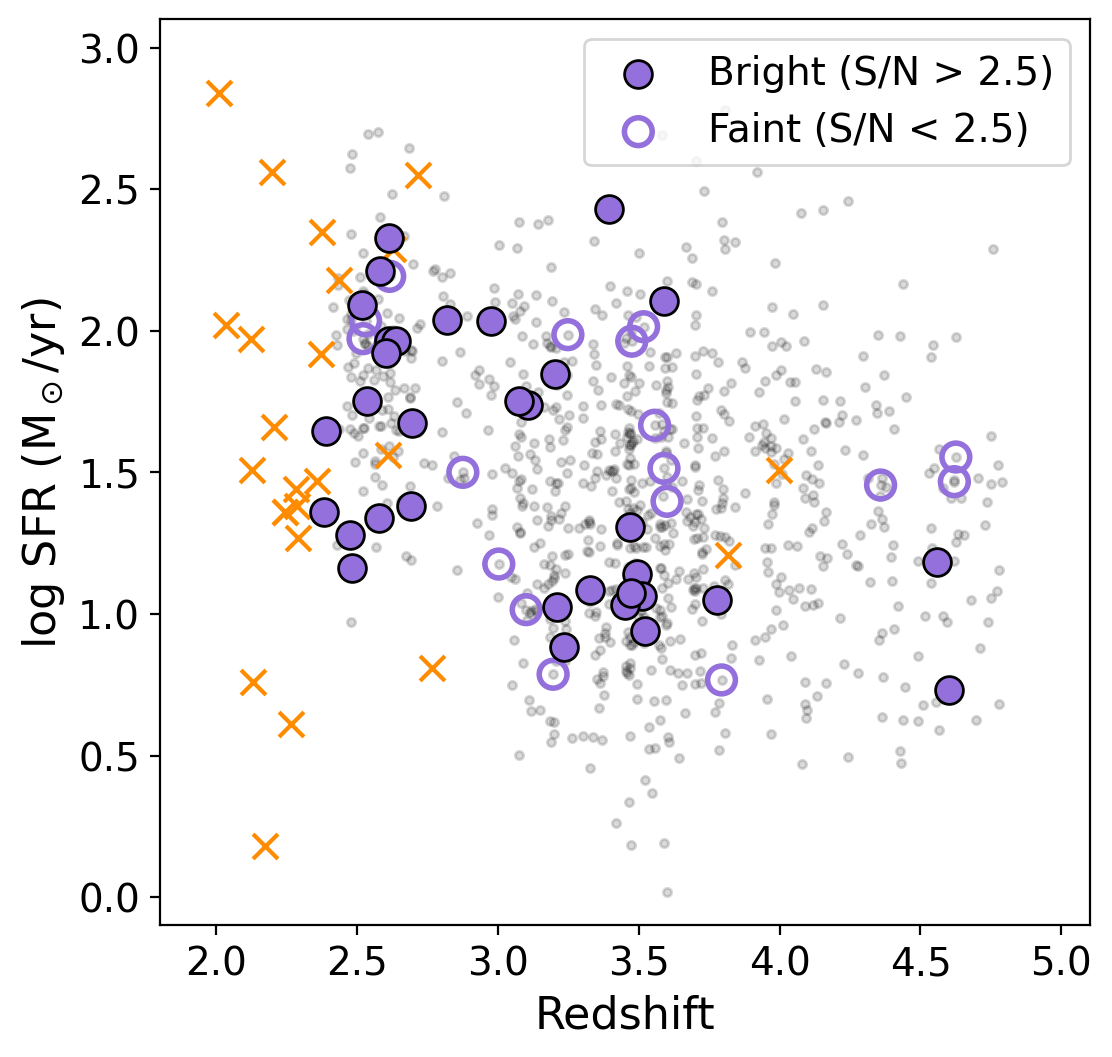}
        \end{minipage}
        \begin{minipage}{0.32\textwidth}
                \centering
                \includegraphics[scale=0.42]{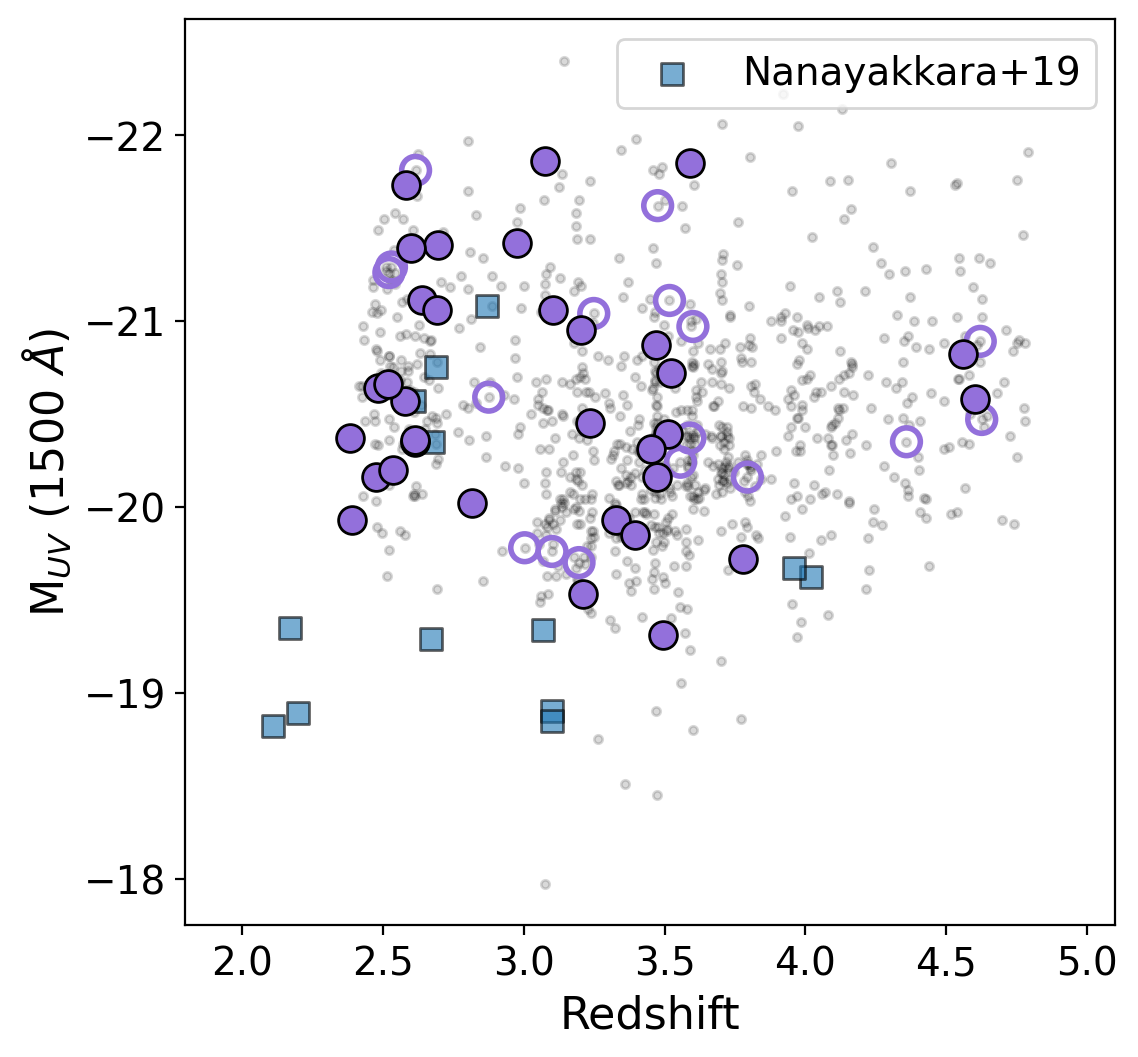}
        \end{minipage}
        \caption{Distribution of the derived stellar masses (\emph{left}), star-formation rates (\emph{centre}), and rest-frame UV magnitudes (\emph{right}) of \emph{Bright} (filled circles) and \emph{Faint} (open circles) \heii emitters with redshift. Also shown are measurements for the Parent sample (grey dots) and for \heii emitters reported by \citet{cas13} (orange crosses) and \citet{nan19} (blue squares). For our sample, we do not see any significant differences between the stellar masses, SFRs, and UV magnitudes of \heii emitters and the parent sample. Our measurements are also comparable with those from the literature, although the \citet{cas13} sample probes lower redshifts compared to both the \citet{nan19} and our \heii emitters.}
        \label{fig:muv-mstar-z}
\end{figure*}

We find that for \emph{Bright} \heii emitters in our sample, the stellar masses range from $ \log_{10} M_\star  = 8.8 - 10.7$ $M_\odot$ and for \emph{Faint} emitters, the stellar masses range from $\log_{10} M_\star  = 8.8 - 10.2$ $M_\odot$. In the left panel of Figure \ref{fig:muv-mstar-z} we show the distribution of the derived stellar masses with redshift for both \emph{Bright} and \emph{Faint} sources in our sample. Also shown in Figure \ref{fig:muv-mstar-z} are the stellar masses and redshifts of the Parent sample. 

It is clear from the figure that both \emph{Bright} and \emph{Faint} \heii-emitting galaxies are evenly distributed across stellar mass and redshift. To quantitatively explore any differences between galaxies that show \heii emission and those that do not, a two-sample Kolmogorov-Smirnov (KS) test of the stellar mass distributions of the \heii emitters and the 899 galaxies with no \heii emission was performed. The null hypothesis is that both samples are drawn from the same population of galaxies. The D-statistic returned by the KS test was 0.17, with a p-value of 0.29, which does not rule out the null hypothesis. Therefore, based on their stellar masses, there is no significant difference in the population of galaxies that show \heii emission compared to those that do not.

For comparison, in Figure \ref{fig:muv-mstar-z} we also show stellar masses from \heii-emitting galaxies in the \citet{cas13} sample. Our \heii emitters have comparable masses to the \citet{cas13} sample, although our sample extends to higher redshifts where the observed stellar masses tend to be lower, which is due to VANDELS source selection. More details about target selection and potential biases can be found in \citet{mcl18}.

\subsection{Star formation rates}
The dust-corrected SFRs of \emph{Bright} \heii emitters are in the range $ \log_{10}(\textrm{SFR}) = 0.7 - 2.3$ $M_\odot \textrm{~yr}^{-1}$ and those of \emph{Faint} emitters are in the range $\log_{10}(\textrm{SFR}) = 0.5 - 1.9$ $M_\odot \textrm{~yr}^{-1}$. We show the distribution of SFRs with redshift in the middle panel of Figure \ref{fig:muv-mstar-z}, with measured SFRs from the parent sample and from \citet{cas13} also shown.

We do not find any significant difference between the distribution of SFRs of \heii-emitting galaxies and that from galaxies with no \heii emission. The distribution of SFRs are also comparable with the \citet{cas13} sample, although there are a few galaxies with very high SFRs ($\sim1000$ $M_\odot \textrm{~yr}^{-1}$) in the \citet{cas13} sample that are not present in our sample. The overall spread of SFRs in our sample is also lower than that in the \citet{cas13} sample. The underlying differences in the SFR estimation through SED fitting between the two samples may add to the uncertainties in these quantities. The stellar masses and star-formation rates for \emph{Bright} \heii emitters are also shown in Table \ref{tab:strong_sources}.

\subsection{Specific SFR}
We also compare the specific star-formation rates (sSFR) for \heii emitters with those of the parent sample. Unsurprisingly, the distribution of sSFR for \heii emitters is not significantly different from that of the parent sample. The median specific SFR for \heii emitters is $\log_{10}(\textrm{sSFR})=-7.7$~yr$^{-1}$ and is comparable to the values derived by \citet{cas13} for their sample, although some of their galaxies show much higher sSFRs. The relatively high sSFR of our \heii emitters suggests that overall, these galaxies are in the process of actively assembling their stellar masses and their stellar ages are low, assuming that they have been constantly forming stars at the same rate. 

\subsection{Rest-frame UV magnitudes}
The rest-frame UV magnitudes (at 1500 $\AA$) for \emph{Bright} sources range from $M_{\textrm{UV}} = -21.9$ to $-19.2$, and for \emph{Faint} sources range from $M_{\textrm{UV}}$ = $-21.8$ to $-19.8$. We show the UV magnitudes with redshift for our sample, along with those from the parent sample and the sample of \heii emitters reported by \citet{nan19}.

We find \heii emission originating from galaxies with a wide range of absolute UV magnitudes, all the way to the faintest limits of the survey. However, we again find no significant difference between the underlying populations of \heii emitters and galaxies with no \heii based on their absolute UV magnitudes. The \citet{nan19} sample for comparison probes fainter UV magnitudes than our sample. 

Overall, we find no significant differences in the examined physical properties of galaxies that show \heii emission and galaxies that do not. This is interesting as the ionisation energies required to produce bright \heii emission are quite high, and why some galaxies show \heii while a large majority do not is not immediately clear from the comparison above. In Section \ref{sec:metallicity_comparison} we compare the metallicities of galaxies with \heii emission and to those of galaxies without. In the following sections, we set some constraints on the underlying ionising properties of \heii -emitting galaxies in a bid to understand the processes that may be driving \heii emission.

\section{Comparison with photo-ionisation models}
\label{sec:models}
In this section we take a deeper look at the underlying ionising mechanisms that power the observed \heii emission. We carry out the analysis both on measurements made on individual spectra of \emph{Bright} \heii emitters, as well as stacks of spectra created in bins of \heii FWHM. We also measure the emission line fluxes of the stack of all the \emph{Faint} \heii emitters (presented in Section 2.3). 
\begin{table*}
        \centering
        \caption{Ultraviolet line measurements for individual galaxies with high enough S/N for the lines.}
        \begin{tabular}{l c c c c c c c c r}
        \hline
         & \multicolumn{3}{c}{\heii $\lambda1640$} & \multicolumn{3}{c}{\oiii $\lambda1661 + \lambda1666$} & \multicolumn{3}{c}{\ciii $\lambda1909$} \\
        ID & Flux & FWHM & EW & Flux & FWHM & EW & Flux & FWHM & EW \\
           & \tiny{(erg~s$^{-1}$~cm$^{-2}$)} & \tiny{(km s$^{-1}$)} & \tiny{($\AA$)} & \tiny{(erg~s$^{-1}$~cm$^{-2}$)} & \tiny{(km s$^{-1}$)} & \tiny{($\AA$)} & \tiny{(erg~s$^{-1}$~cm$^{-2}$)} & \tiny{(km s$^{-1}$)} & \tiny{($\AA$)} \\
        \hline \hline
        
        \emph{Narrow} \\
    \hline
  
        CDFS015374 & 1.1 & 240 & 1.7 & 1.6 $\pm$ 1.4 & 250 $\pm$ 150 & 3.2 $\pm$ 2.8 & 3.1 $\pm$ 2.4 & 510 $\pm$ 300 & 6.5 $\pm$ 5.0 \\
        CDFS023170 & 2.8 & 455 & 1.1 & 1.3 $\pm$ 1.1 & 315 $\pm$ 300 & 0.7 $\pm$ 0.6 & 6.5 $\pm$ 2.4 & 850 $\pm$ 150 & 3.2 $\pm$ 0.4 \\
        CDFS023527 & 1.5 & 260 & 0.9 & 2.6 $\pm$ 1.4 & 270 $\pm$ 110 & 1.6 $\pm$ 0.9 & 8.7 $\pm$ 3.1 & 580 $\pm$ 110 & 6.9 $\pm$ 2.4 \\
        CDFS113062 & 4.0 & 640 & 1.4 & 2.4 $\pm$ 1.8 & 450 $\pm$ 225 & 0.9 $\pm$ 0.7 & 12.3 $\pm$ 2.2 & 630 $\pm$ 50 & 5.6 $\pm$ 1.0 \\
        CDFS126819 & 9.5 & 860 & 8.0 & 2.4 $\pm$ 1.8 & 220 $\pm$ 170 & 2.4 $\pm$ 1.8 & 9.9 $\pm$ 4.0 & 820 $\pm$ 150 & 8.0 $\pm$ 3.2 \\
        UDS013586 & 3.1 & 335 & 0.6 & 2.8 $\pm$ 2.4 & 400 $\pm$130 & 0.6 $\pm$ 0.5 & 12.0 $\pm$ 3.1 & 800 $\pm$ 100 & 3.3 $\pm$ 0.8 \\ 
        UDS019505 & 2.0 & 530 & 0.7 & 2.2 $\pm$ 2.0 & 420 $\pm$ 140 & 0.9 $\pm$ 0.8 & 8.3 $\pm$ 2.4 & 860 $\pm$ 120 & 4.5 $\pm$ 1.3 \\
        UDS281893 & 2.9 & 660 & 1.1 & 2.1 $\pm$ 1.1 & 450 $\pm$ 70 & 0.9 $\pm$ 0.5 & 7.1 $\pm$ 1.3 & 600 $\pm$ 60 & 4.1 $\pm$ 0.2 \\
        CDFS009705 & 2.2 & 510 & 1.3 & - & - & - & 8.6 $\pm$ 1.1 & 820 $\pm$ 50 & 6.3 $\pm$ 0.8 \\
        CDFS229681 & 1.2 & 520 & 1.9 & - & - & - & 3.9 $\pm$ 2.7 & 600 $\pm$ 290 & 8.6 $\pm$ 6.0 \\
        
        \hline
        \emph{Broad} \\
    \hline
        UDS137388 & 23.1 & 1420 & 4.6 & - & - & - & 16.0 $\pm$ 6.3 & 1100 $\pm$ 280 & 5.0 $\pm$ 2.0 \\
        \hline
        \end{tabular}
        \label{tab:uvlines}     
        \tablefoot{Line fluxes are in units of $10^{-18}$ \flux. The \oiii FWHM given in the table is for the $\lambda1666$ line. The errors on \heii line properties are given in Table \ref{tab:strong_sources}.}
\end{table*}

To compare UV emission line ratios with models, we primarily rely on the \heii, \oiii and \ciii lines given the line strengths and the wavelength coverage based on the redshift distribution of sources. We include the analysis of UV line ratios of individual sources for which the S/N of each of these emission lines is greater than 2.5, and in Table \ref{tab:uvlines} we give the measured UV emission line properties for these sources.

\subsection{Comparison with \citet{gut16}}
\begin{figure*}
        \centering
        \begin{minipage}{0.45\textwidth}
                \includegraphics[scale=0.6]{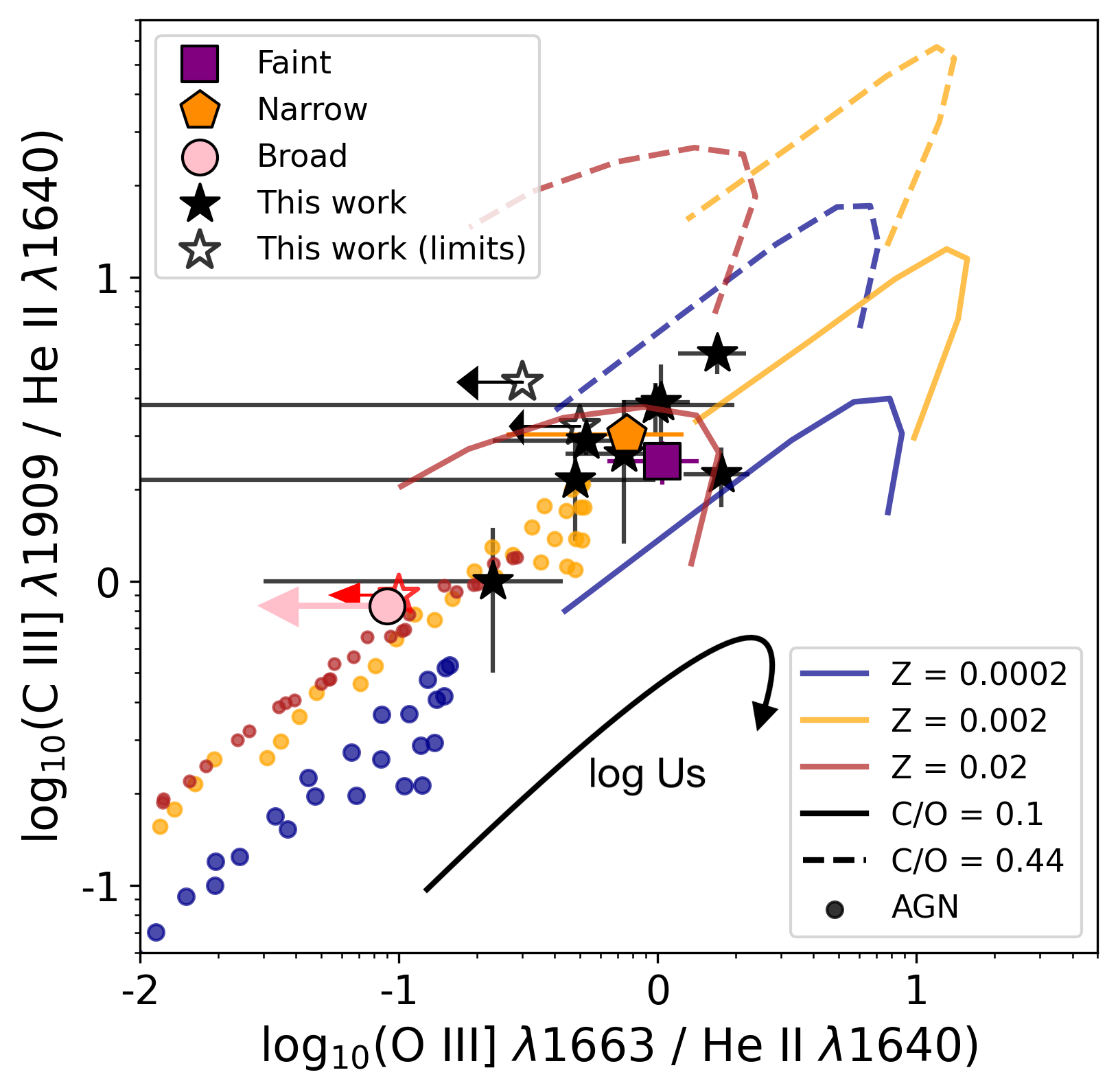}
        \end{minipage}
        \begin{minipage}{0.45\textwidth}
                \includegraphics[scale=0.6]{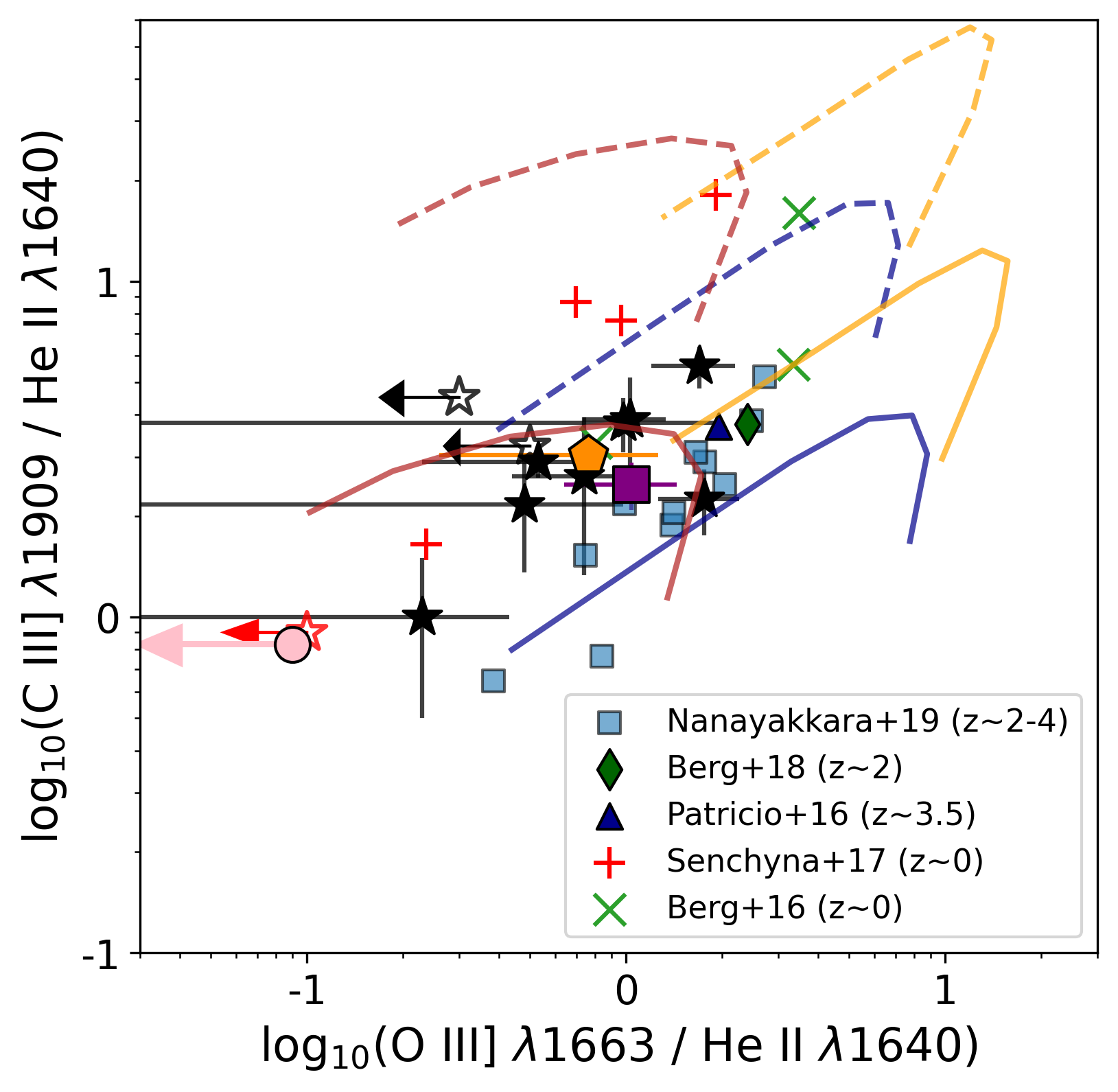}
        \end{minipage}
        \caption{\textit{Left:} UV line ratios determined for individual galaxies (stars, where empty stars are limits) and stacks of \heii emitters (coloured symbols), compared with photo-ionisation models from star-forming galaxies (lines) by \citet{gut16} and AGNs (dots) by \citet{fel16}. The red star indicates limits on a broad \heii -emitting galaxy. Both models for star-formation and AGNs are shown for three metallicities: $Z = 0.0002, 0.002$ and $0.02$. For star-formation models, the solid lines indicate C/O = 0.1, and the dashed lines indicate C/O = 0.44 (solar value). \textit{Right:} Zoom in on the star-forming region with AGN models removed for clarity. We also show line ratios from the sample of \heii emitters of \citet{nan19}. Also shown are measurements from local low-metallicity galaxies from \citet{sen17} and \citet{ber16}, and individual lensed galaxies at $z=2.2$ \citep{ber18} and $z=3.5$ \citep{pat16}. Line ratios of individually detected galaxies from our sample, as well as stacks of \emph{Faint} and \emph{Narrow} \heii emitters favour the star-forming models, whereas the stack of \emph{Broad} emitters occupies a region between the star-forming and AGN models. The separation between narrow and broad emitters suggests that they are probably powered by different ionising mechanisms and/or conditions.}
        \label{fig:gutkin_feltre}
\end{figure*}
We now compare our results with the predictions from photo-ionisation models for nebular emission from star-forming galaxies from \citet{gut16}. The models make use of the latest version of the stellar population synthesis code by \citet{bc03} in combination with the photo-ionisation code \textsc{Cloudy} \citep{fer13}. The key parameters dictating the line ratios obtained from their models are the metallicity of the ISM, the ionisation parameter, the dust-to-metal mass ratio, the carbon-to-oxygen abundance ratio (C/O), the gas density of hydrogen, and the upper mass cutoff of the stellar IMF. For the purposes of our analysis, we use three different ISM metallicities: $Z = 0.0002, 0.002,$ and $0.02$, where the solar metallicity is $Z_\odot \approx 0.02$. We also use two different C/O values for each metallicity: C/O = 0.1 and 0.44, where solar C/O $= 0.44$ \citep{gut16}. The C/O values observed for star-forming galaxies at $z\sim3$ have been seen to lie within this range \citep[see e.g.][]{amo17}. We fix the dust-to-metal mass ratio at 0.3, which is the present-day Galactic value, and the hydrogen gas density to 100 cm$^{-3}$, as this is the closest value available in the models to the measured typical value of 250 cm$^{-3}$ for gas density in star-forming galaxies at high redshifts \citep{san18}. We find that there is no difference in the model outputs for $n_H=100$ cm$^{-3}$ or 250 cm$^{-3}$. We use an upper mass cutoff of 300 $M_\odot$ so as to capture the contribution from the most massive stars. The predicted line ratios from the model are a result of constant star formation at a rate of 1 M$_\odot$ yr$^{-1}$ for 100 Myr. The dimensionless ionisation parameter ($U_s$), which is the ratio of the ionising photon flux and the gas density \citep[see][for a detailed explanation]{kew19}, is left free for each track.

This analysis is focused on the diagnostic that uses the line ratios \oiii/\heii versus \ciii/\heii. For \oiii flux, we sum the contributions of the \oiii $\lambda1661$ and $\lambda1666$ lines. There are eight galaxies in our sample that have enough SNR for all three lines to enable a reliable individual line ratio measurement. Additionally, there are three \heii -emitting galaxies with detections of \ciii only, and for these we use limits on \oiii. We also analyse the line ratios determined from the stacked spectra of the different classes of \heii emitters mentioned earlier. In Figure \ref{fig:gutkin_feltre} we show the measured line ratios of individual sources and stacks, with the star-formation models from \citet{gut16} ranging in ionisation parameter values, $-4< \log U_s < -1,$ and AGN models from \citet{fel16} ranging within $-5 < \log U_s < -1$. The AGN models have a gas density $n_\textrm{H} = 1000$ cm$^{-3}$ for AGNs with a dust-to-metal ratio of $\xi_\textrm{d} = 0.3$. We find that using other (fiducial) values of $n_\textrm{H}$ or $\xi_\textrm{d}$ does not have a major impact on the parameter space occupied by AGNs. Models for three different metallicities are used: $Z=0.0002, 0.002,$ and $0.02.$ The value of the ionisation parameter for each model ($\log U_s$) increases from the bottom left towards the top right in the Figure \ref{fig:gutkin_feltre}.

Given the weak line strengths of other UV lines observed in individual galaxy spectra, a quantitative analysis performed using chi-squared minimisation is beyond the scope of this work. However, we provide some qualitative constraints on the ionising source properties of \heii -emitting galaxies based on the model predictions. Additionally, to put into context the properties of our sample of \heii emitters, we also compare our line ratios with galaxies that have similar properties in the literature. Figure \ref{fig:gutkin_feltre} shows that the UV line ratios of the majority of individual \heii emitters favour ionisation from star formation as opposed to AGNs. There are degeneracies between the model metallicities that can reproduce the line ratios. The subsolar metallicities with C/O ratios between $0.1$ and $0.38$ require lower ionisation parameter values in the range  $-3 < \log U_s < -2$ to reproduce the line ratios, whereas solar metallicity models require $\log U_s > -2$ to reproduce the observed line ratios. This degeneracy is primarily linked to the trade-off between the shape of the ionising radiation field and the number of ionising photons produced. Compared to higher metallicity populations, lower metallicities give rise to harder ionisation fields, which means that comparatively lower ionisation parameter values can still produce enough energetic ionised photons to deposit the required amount of total kinetic energy in the gas to explain the line ratios that we see in our sample.

The line ratios from stacks of both \emph{Faint} and \emph{Narrow} \heii emitters overlap with those observed in individual galaxies with narrow \heii favouring ionisation by star formation. However, the line ratios from the \emph{Broad} stack  are hard to explain using star-formation alone and favour photo-ionisation by AGNs with solar to mildly subsolar metallicities. The \oiii line strength in the stack of \emph{Broad} emitters is weak, and therefore we can only set a limit on the \oiii/\heii ratio. The clear separation between the \emph{Narrow} and \emph{Broad} \heii emitters illustrates the possible differences in ionising mechanisms that power the \heii emission in each case.

Figure \ref{fig:gutkin_feltre} also shows line ratios from the sample of \heii emitters from MUSE presented by \citet{nan19}; this sample overlaps well with  our sample in terms of redshift distribution ($z\sim2-4.5$) and offers the best sample for a direct comparison. We also show line ratios from individual low-mass, low-metallicity lensed galaxies at $z=2.2$ \citep{ber18} and at $z=3.5$ \citep{pat16} that show extreme UV-ionising spectra, representing the conditions likely to be prevalent in galaxies that reionised the Universe at $z>6$. There is considerable overlap between the line ratios of our sample and those determined by \citet{nan19}, and measurements from lensed galaxies at $z\sim2-3$ are also similar to those seen in our \heii emitters.

Additionally, we show line ratios observed in local metal-poor \heii-emitting galaxies from \citet{sen17} and \citet{ber16}. From both these local samples we select only those sources with reliable measurements of all three emission lines in question.  The $z\sim0$ low-metallicity galaxies have slightly higher \ciii/\heii ratios, but comparable \oiii/\heii ratios to our sample, which may be explained by higher C/O ratios in the local Universe compared to high redshifts. Line ratios of $z\sim0$ galaxies considered in this study also tend to favour higher stellar metallicities, which could explain the higher C/O ratios in local galaxies as C/O for star-forming galaxies has been found to be linked to the O/H ratio \citep{amo17}. However, there are degeneracies between metallicity and C/O from the models, and the dominant effect that results in slightly differing line ratios between local galaxies and our sample is not entirely clear.

\subsection{Comparison with \citet{xia18} -- Binary stellar population}
We now compare the UV line ratios of our galaxies and stacks with predictions from the Binary Population and Spectral Synthesis Code (BPASS; \citealt{eld17, sta18}), which are presented in \citet{xia18}. Since very high effective temperatures and hard ionising spectra in galaxies are required to power the nebular \heii emission line, the inclusion of interacting binary stars in stellar population synthesis models may hold the key \citep{eld17, got18, xia18, got19}. The inclusion of binary evolution in the modelling of stellar populations allows the contribution of a harder UV-ionising spectrum with a longer duty cycle to be captured \citep{eld17}. 

\begin{figure}
        \centering
        \includegraphics[scale=0.6]{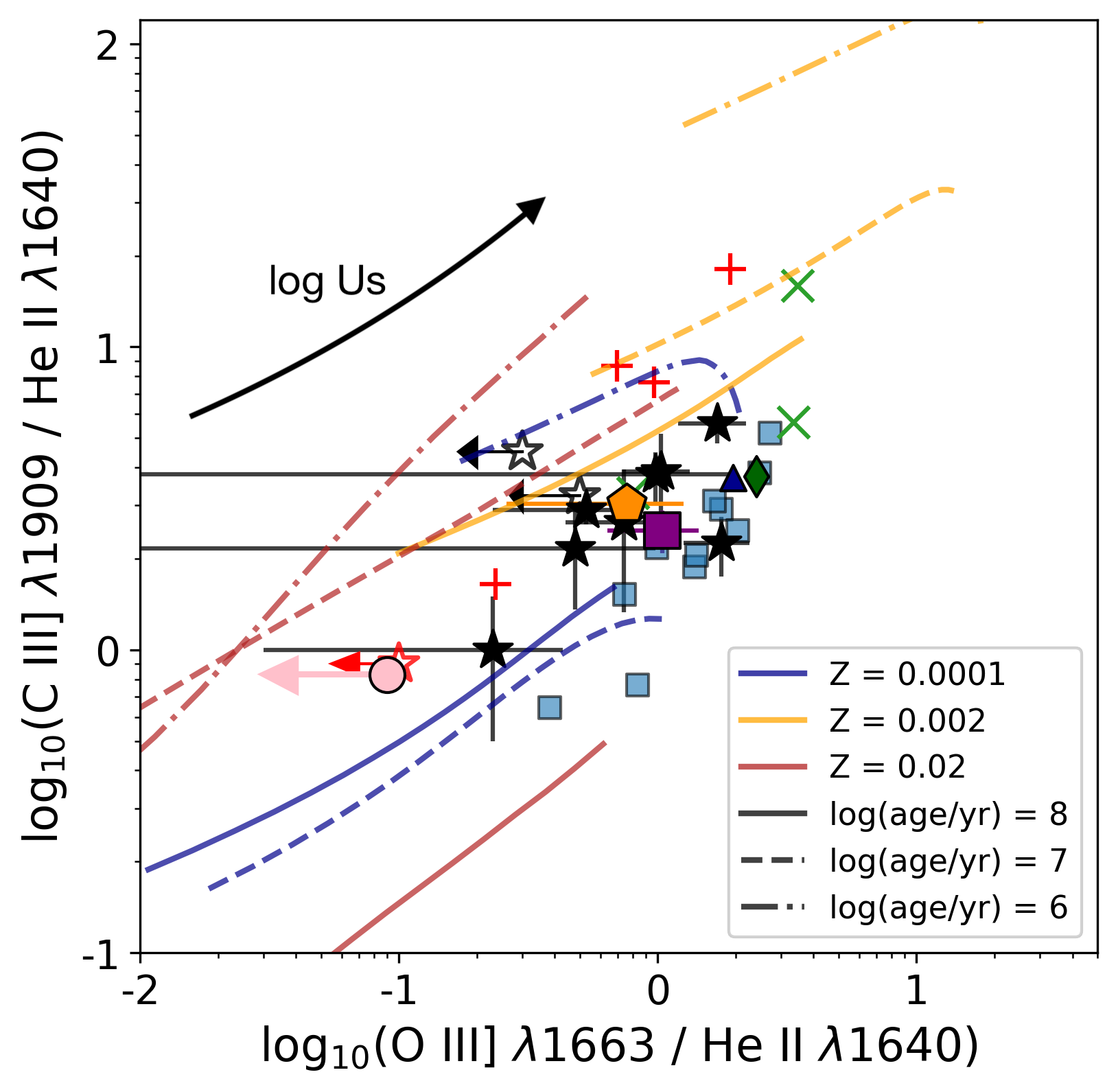}
        \caption{UV line ratios determined for individual galaxies and stacks of \heii emitters. The symbols are the same as in Figure \ref{fig:gutkin_feltre}. The line ratios are compared with photo-ionisation models from star-forming galaxies including binary stars from \citet{xia18} with three metallicities, $Z = 0.0001, 0.002,$ and $0.02$, and stellar ages $10^6$ (dot-dashed), $10^7$ (dashed), and $10^8$ (solid) years. Also shown are line ratios from \citet{nan19}, \citet{sen17}, \citet{ber16}, \citet{ber18}, and \citet{pat16} for comparison, represented by the same coloured symbols as in Figure 6. The majority of individual detections and stacks of \emph{Narrow} and \emph{Faint} \heii emitters favour subsolar metallicities and $\log U_s > -2$, whereas the \emph{Broad} emitters favour solar metallicities and lower $\log U_s$ values.}
        \label{fig:xiao}
\end{figure}

For comparison with the predicted UV line ratios, we once again use \oiii/ \heii versus \ciii/ \heii from the \citet{xia18} models, analysing individual galaxy spectra as well as stacks. We also again set the hydrogen gas density, $n_H$ , to 100 cm$^{-2}$, and choose ISM metallicities of $Z = 0.0001, 0.002,$ and $0.02$. The \citet{xia18} models evolve as a single instantaneous starburst with ages varying from 1 Myr to 10 Gyr, and for this analysis we select models with stellar ages, $10^6$, $10^7$ , and $10^8$ years. The C/O ratio in these models is fixed to the solar value of C/O = 0.44. The resulting line ratios predicted by the models as well as those measured in individual galaxies and stacks are shown in Figure \ref{fig:xiao}. For each model, the ionisation parameter ranges $-3.5 < \log U_s < -1.5$ and increases from the bottom left to the top right of the figure, as indicated by the black arrow.

The binary models from \citet{xia18} seem to fit the observed line ratios more consistently with respect to degeneracies within model predictions of different metallicities, both for individual galaxies and stacks. For example, the majority of individual detections as well as the \emph{Faint} and \emph{Narrow} stacks favour subsolar metallicities in the range $Z = 0.0001 - 0.002$, with ionisation parameter $\log U_s > -2$ and stellar ages of $10^8$ years. Interestingly, binary stellar models suggest that the line ratios observed in \emph{Broad} \heii emitters can be explained purely by star formation and favour higher metallicities compared to the narrow emitters. A star-formation explanation is in contrast to the predictions from single-star models from \citet{gut16}, where additional ionising photons from AGNs were required to explain the line ratios. The fact that \emph{Broad} \heii emitters can be explained by higher metallicity models is consistent with the picture where WR stars, predominantly formed at higher metallicities, are the dominant sources of \emph{Broad} \heii seen in spectra of star-forming galaxies. Due to binary interactions implemented in the \citet{xia18} models, stars can spend a longer amount of time in the WR phase, thereby providing the required \heii ionising photons for prolonged periods \citep{eld17}. 

As a more direct test of whether the BPASS models can fully account for the number of He$^+$ photons produced in the sources presented in this paper, we now compare the measured and predicted EWs of the three UV emission lines in this section.
\begin{figure*}
        \centering
        \begin{minipage}{0.45\textwidth}
                \includegraphics[scale=0.6]{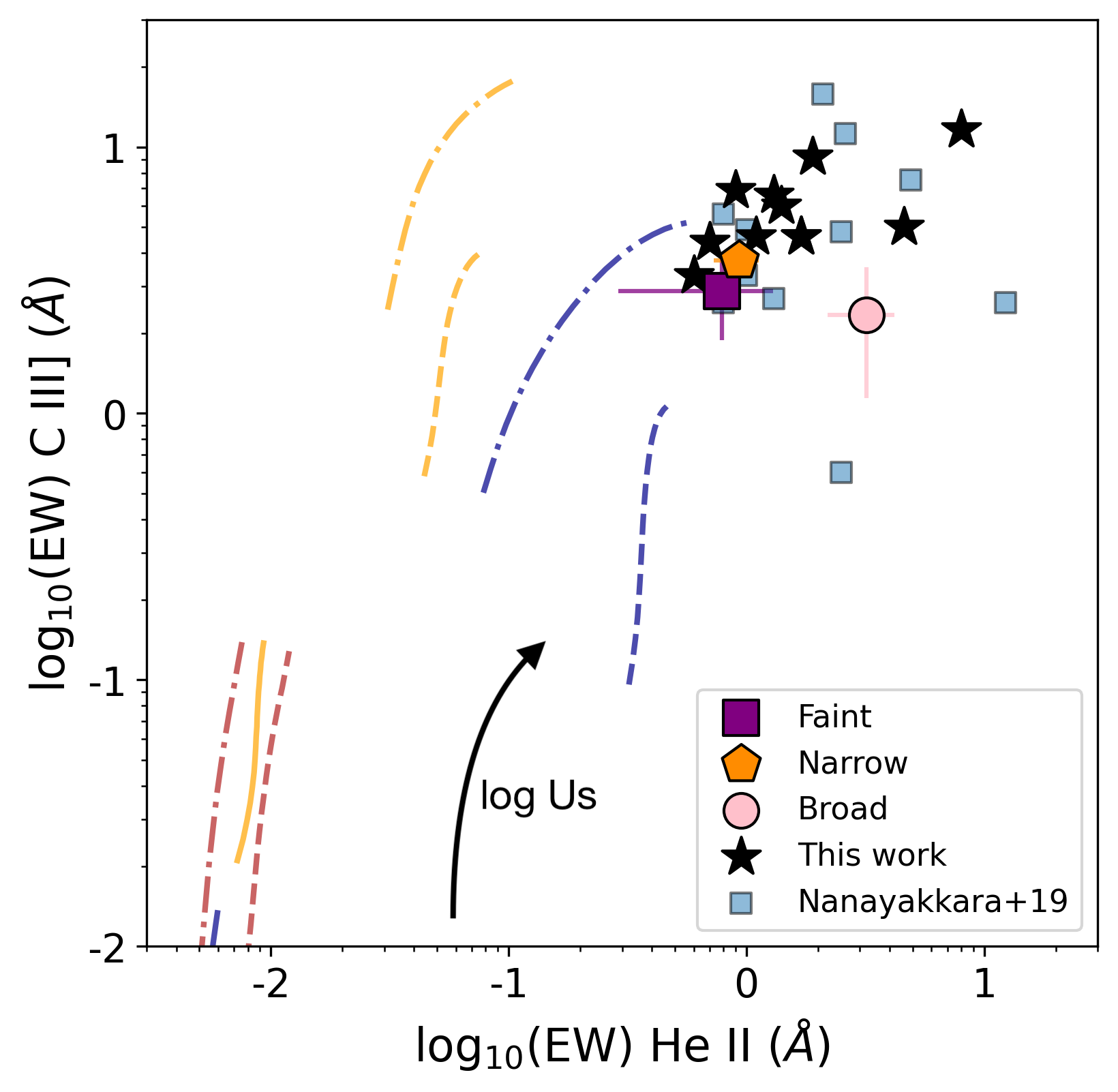}
        \end{minipage}
        \begin{minipage}{0.45\textwidth}
                \includegraphics[scale=0.6]{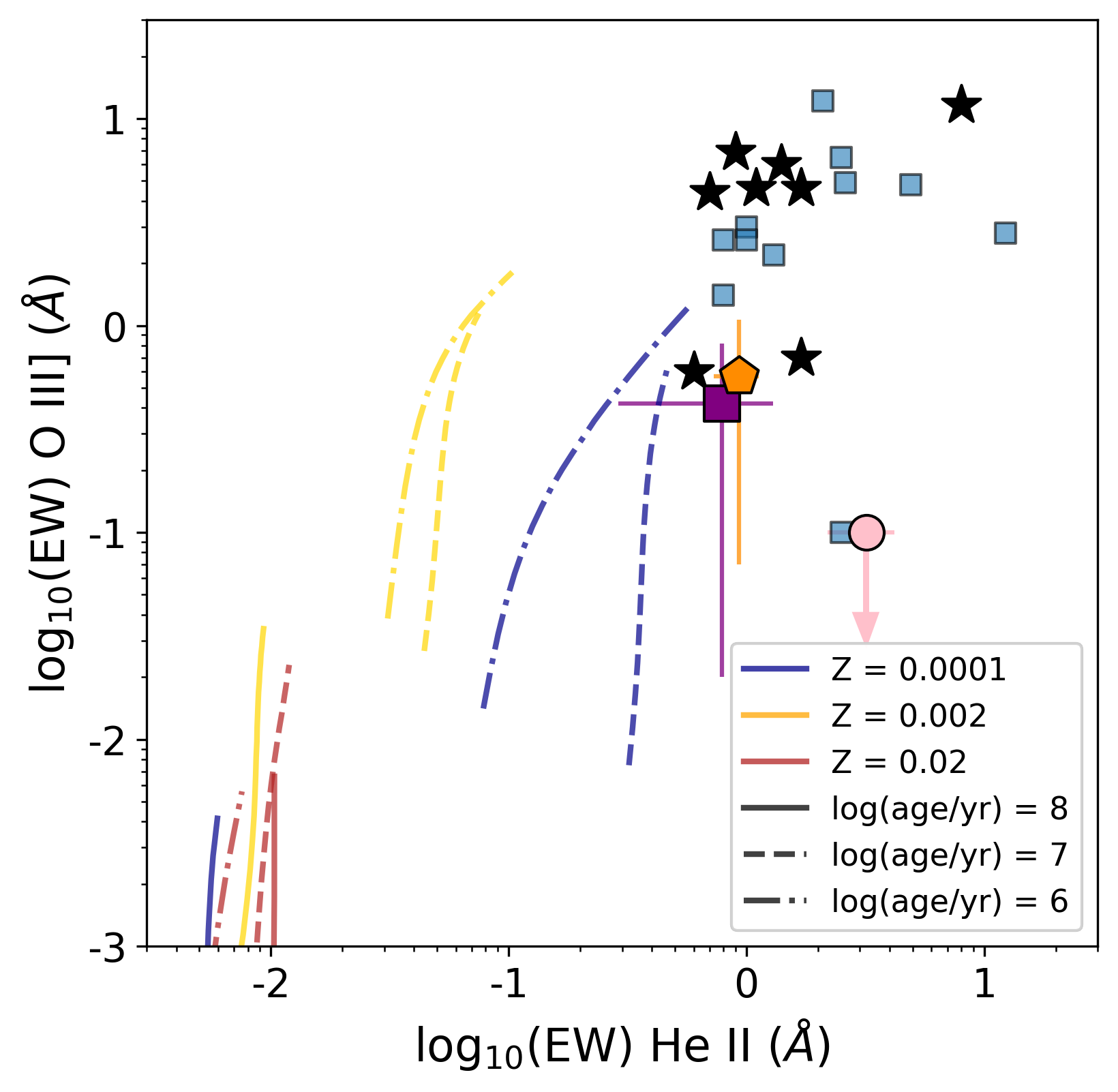}
        \end{minipage}
        \caption{Comparison of EWs measured in individual galaxies (black stars) and stacks (coloured symbols) with predictions from BPASS models. The left panel shows EWs of \heii and \ciii, and the right panel shows EWs of \heii and \oiii . Also shown are EW measurements from \citet{nan19}. The measurements from individual sources in our sample are in close agreement with those of \citet{nan19}. The BPASS models underpredict the \heii EW, but subsolar metallicity models can reproduce the \ciii EW. The \oiii EWs are also underpredicted at all metallicities for the majority of sources. Solar metallicity models underpredict the EWs by several orders of magnitude, and therefore we can conclude that based on the observed UV line EWs, \heii -emitting galaxies tend to favour subsolar metallicities. To properly account for the missing \heii ionising photons, additional mechanisms such as stripped stars \citep{got18, got19} or X-ray binaries \citep{sch19} may be needed.}
        \label{fig:xiao-ew}
\end{figure*}

In Figure \ref{fig:xiao-ew}. we show the distribution of \heii , \ciii , and \oiii  EWs from individual measurements as well as stacks, along with the predictions from \citet{xia18} models for the same ages and parameters as before. We note that for individual sources, we only show the detections and not limits on measured EWs. Also shown for comparison are EW measurements from \citet{nan19}. We find that \heii emitters in our sample have similar \heii, \oiii, and \ciii EWs to those in the \citep{nan19} sample. Overall, the stacks show lower \oiii EWs compared to individual detections, which is not surprising as the stacks contain multiple sources that do not show strong \oiii.

The \citet{xia18} models with a metallicity of $Z=0.002$ and low stellar ages ($10^6-10^7$ years) can reproduce the observed \ciii EWs, which is consistent with the photo-ionisation modelling from \citet{nak18}, but they fall short of reproducing the \heii EWs for both individual sources and stacks. We find that the highest \heii EWs are predicted by the lowest metallicity ($Z=0.0001$) models with stellar ages in the range $10^6 - 10^7$ years, but at this metallicity the observed \ciii EWs are only reproduced by models with a stellar age of $\sim10^6$ years. However, these models  cannot reproduce the observed \oiii EWs for individual sources, but the EWs measured in the stacked spectra can be explained by models with metallicities $Z=0.0001-0.002$. Solar metallicity models ($Z=0.02$) underpredict the EWs for all the UV lines considered by more than two orders of magnitude, effectively ruling out a solar metallicity stellar population origin of these lines. 

Therefore, we conclude that the BPASS models can reproduce the observed UV line ratios but not line strengths, similar to what was reported by \citet{nan19}. These latter authors showed that even after calibrating for the \ciii deficit between observations and models, the BPASS models still underpredict the number of \heii ionising photons by almost an order of magnitude. However, the very low metallicity models from BPASS come close to reproducing the observed \heii EWs after applying the calibration. From our results and those of \citet{nan19}, it is clear that  additional sources that produce \heii ionising photons are required to properly account for the observations. Recently, stripped stars \citep{got18, got19} and X-ray binaries \citep{sch19} were suggested as extra contributors of \heii ionising photons that could potentially bridge the gap left by stellar population synthesis models in order to reproduce the \heii line strengths. We explore these possibilities in more detail in the following section.

\subsection{Revisiting the \heii emitters with strong \civ}
\label{sec:civ_emitters}
In the initial classification of sources, we used the presence of \civ emission in galaxy spectra as an indicator of AGN activity. The ionising potential required for the \civ emission line is very high ($\sim49.9$ eV) and close to the \heii ionising potential. Therefore, standard stellar population models based on single stars are unable to reproduce strong \civ lines without including a contribution from AGNs. However,  low-metallicity binary star models are capable of producing \heii ionising photons (and therefore, \civ ionising photons) for longer periods of time, and therefore we revisit the spectra of sources that show \civ emission and use the BPASS models to probe their underlying ionising mechanisms. 

\begin{figure}
        \centering
        \includegraphics[scale=0.6]{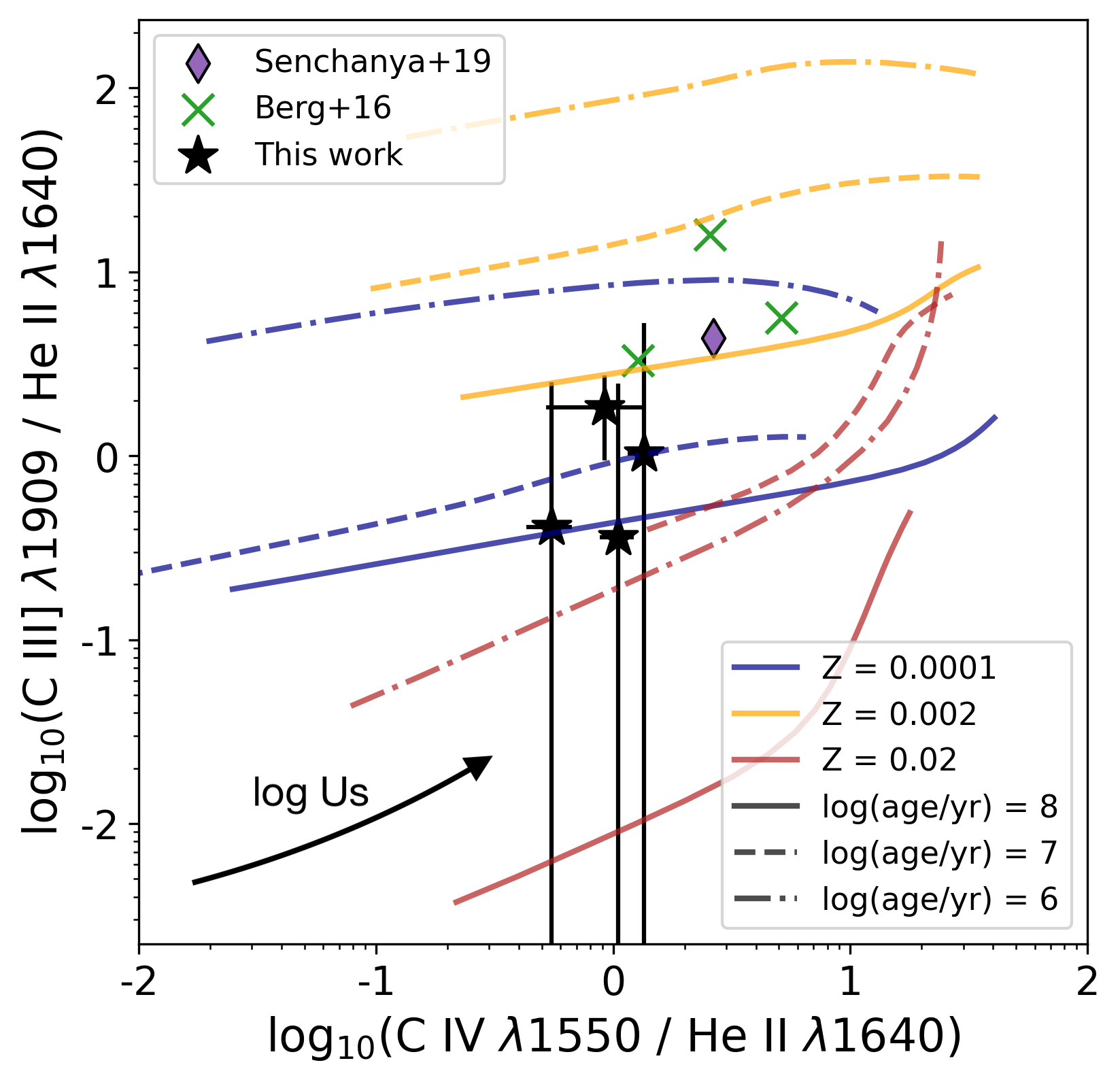}
        \caption{Line ratios of \heii -emitting galaxies that show strong \civ emission in their spectrum. Model predictions from BPASS are shown, which are able to reproduce the observed line ratios using star-formation activity alone. Also shown for comparison are line ratios measured in local low-metallicity galaxies by \citet{ber16} and \citet{sen19}. When compared to the local galaxies, our sources at higher redshifts favour lower metallicities.}
        \label{fig:Xiao-CIV}
\end{figure}

In this analysis, we only select those galaxies that do not have an X-ray counterpart in the available catalogues and are therefore not clearly AGNs. Using BPASS model predictions, we find that the line ratios \civ/ \heii versus \ciii/ \heii observed for our \heii emitting galaxies with strong \civ emission can be explained using star-formation activity alone, as shown in Figure \ref{fig:Xiao-CIV}. The best-fitting models are those with subsolar metallicities ($0.0001-0.002$) and stellar ages in the range $10^7-10^8$ yr. We note that one \civ emitter does not show any \ciii emission, and therefore a higher metallicity for this particular source cannot be categorically ruled out based on the model predictions. The models shown here are the same as Figure \ref{fig:xiao}. We note that this analysis does not include the source UDS021234 as the \ciii line is not covered in the observed spectrum because of its high redshift.

Several other studies have reported the detection of both \civ and \heii emission in low-metallicity galaxies, predominantly in the local Universe. In the \citet{ber16} sample of local dwarf galaxies, three \heii emitters show strong double-peaked \civ emission. Recently, \citet{sen19} found one of the most prominent nebular \civ doublets in a local extremely low-metallicity galaxy, and show that \civ emission may be ubiquitous in extremely metal-poor galaxies with very high specific star formation rates. This is also expected, for example, from photo-ionisation models of \citet{nak18}. \citet{ber19} also reported the detection of the \civ doublet from two $z\sim0$ \heii -emitting galaxies, and show that the presence of \civ emission may indicate both the production and transmission of very high-energy ionising photons. The common property observed in galaxies that show both \heii and \civ emission is that they tend to be low-metallicity star-forming galaxies, which is also apparent from our sources. 

The presence of both \heii and the resonantly scattered \civ emission lines suggests that a significant number of high-energy ionising photons with energies greater than $\sim49.9$ eV are being produced and transmitted \citep{ber19}. The escape of such high-energy photons in significant amounts is needed from low-mass star-forming galaxies at $z>6$ to drive the process of reionisation, and galaxies with similar properties at intermediate redshifts ($z\sim2-5$) as well as dwarf galaxies in the local Universe can serve as analogues to explore the physical conditions prevalent in galaxies that are expected to reionise the Universe at early times.

\subsection{Effects of different stellar population parameters}
In this section we briefly outline what impact is seen on the line ratios and inferred physical properties by changing certain adjustable parameters that go into the stellar population synthesis modelling.

\vspace{6pt}
\textit{\textbf{(i) Star-formation history}}

An important difference between the \citet{gut16} single-star models and \citet{xia18} binary-star models is the assumed star formation history (SFH). The \citet{gut16} models assume a constant star formation rate of 1 M$_\odot$~yr$^{-1}$ for 100 Myr, whereas the \citet{xia18} BPASS models assume instantaneous star formation at a given time in the galaxy's evolutionary history. As \citet{gut16} note, in their single-star models, most ionising photons are released during the first 10 Myr of stellar evolution. This is a relatively short timescale compared to the lifetime of a galaxy and therefore it is close to quasi-instantaneous star formation, similar to what is assumed in the implementation of the BPASS models. 

\citet{eld17} showed that at low stellar ages, the BPASS binary-star models produce very similar outputs to the \citet{bc03} model on which the \citet{gut16} models are based. The difference between the two models is especially not large at rest-frame UV wavelengths, where the light is dominated by young stars. The difference between star-formation histories is larger at redder wavelengths, where the SED is dominated by older stars. Since we are primarily interested in the UV line ratio predictions, the differences arising from assumed star-formation histories are kept to a minimum. In the context of this study, if the BPASS models also assumed a constant star formation rate for a set period of time, the galaxy ages that best fit the UV line ratios may be higher than what an instantaneous star formation history would suggest.

\vspace{6pt}
\textit{\textbf{(ii) Carbon-to-oxygen abundance ratio (C/O)}}

For the single star models used in this analysis, we show model outputs for two C/O ratios: 0.44 (solar) and 0.1 (sub-solar). For the binary models, the C/O is fixed to the solar value of 0.44. As \citet{gut16} showed, and as is also visible from the model outputs shown in Figure \ref{fig:gutkin_feltre}, the greatest impact of choosing a higher C/O value is an increase in the predicted \ciii / \heii, \ciii / \oiii, and \civ / \heii ratios. \citet{amo17} showed that the C/O ratios in star-forming galaxies at $z\sim2-3$ are subsolar. This means that implementing subsolar C/O ratios in the \citet{xia18} binary models would decrease the predicted \ciii / \heii ratios in Figure \ref{fig:xiao}. This would affect the stellar age of models that would then best fit the observed UV line ratios. For example, for models with a stellar metallicity of $Z=0.002$, younger stellar ages would be able to explain the observed line ratios in our sample of \heii emitters.

\vspace{6pt}
\textit{\textbf{(iii) Dust-to-metal mass ratio ($\mathrm{\xi_d}$)}}

Studies show that $\xi_d$ is expected to vary with redshift, and is closely related to the ISM metallicity of a galaxy. It is expected that at higher redshifts and at lower ISM metallicities, $\xi_d$ decreases \citep{ino03, qi19}. As dust grains are made from metals injected into the ISM by stars, lower metallicities result in less dust production. Therefore, $\xi_d$  can have an additional impact on the observed emission line ratios predicted by models. However, for simplicity, most studies generally assume a fixed $\xi_d = 0.3$, the Galactic value.

\citet{gut16} showed that the choice of $\xi_d$ from 0.1 to 0.5 at lower ISM metallicities has a negligible impact on the \ciii / \heii, \oiii/\heii , and \ciii/\oiii ratios primarily used in this paper. The difference is also relatively minor at solar metallicity, the highest metallicity considered in this study. Therefore, we can conclude that for the metallicities considered in this paper, variation in the value of $\xi_d$ does not result in significant differences in the predicted UV line ratios.

\section{Discussion}
\label{sec:discussion}
In this section we put the \heii emitting galaxies presented in this paper in the context of the general galaxy population at these redshifts. In particular, we look at the physical conditions that lead to production of \heii in emission. We also present possible scenarios that may lead to the emergence of strong \heii at high redshifts, and what these observations mean for galaxies at even higher redshifts, a regime that the \emph{James Webb Space Telescope} (JWST) will probe. 

\subsection{Do \heii -emitting galaxies have systematically lower metallicities?}
\label{sec:metallicity_comparison}
One of the requirements for \heii emission from star-forming galaxies at any redshift is the presence of a young \citep{ber97}, low-metallicity (subsolar) stellar population that is capable of producing enough \heii ionising photons \citep[see e.g.][]{sch03}. Ultraviolet spectra of several low-metallicity dwarf galaxies in the local Universe show strong \heii $1640$ $\AA$ emission \citep{ber16, ber19, sen17, sen19}. \citet{nan19} showed that the UV line ratios of their sample of \heii emitters at $z\sim2-4$ are also best explained by subsolar metallicity models.

Using the predictions of line ratios and the EWs of several UV emission lines from the BPASS models presented in the previous section, it is clear that our \heii -emitting sources also favour subsolar metallicity (Figures \ref{fig:xiao} and \ref{fig:xiao-ew}). Although there are degeneracies between the different parameters of the models, it is clear that solar metallicities may be ruled out. 

To quantitatively measure stellar metallicities of \heii -emitting galaxies, we must rely on stacks that boost the S/N of key features in the spectra. To ensure sufficiently high S/N, we stack a total of 43 sources, previously classified as both `Bright' and `Faint' \heii emitters (removing possible AGNs). The metallicities are then measured following the method of \citet{cul19}, which relies on a combination of stellar population synthesis modelling and metallicity indicators in the UV. With the low S/N of individual galaxy spectra, accurate metallicities using this method can only be obtained for stacks of objects. A slight caveat of the stellar metallicity measurement method from \citet{cul19} used in this paper is that the SED fitting assumes a constant star-formation history. This assumption is valid for averaging across the general star-forming galaxy population, but if galaxies with \heii emission are very young, then their true metallicities may be higher than what is inferred using this method.

For the stacked spectrum of all \heii emitters, we measure a stellar metallicity, $Z=0.0022 \pm 0.0003$, which is roughly 10\% of solar metallicity. We note that the uncertainties are likely to be underestimated due to other systematic errors that are not taken into account. The metallicity is consistent with what the UV line ratio diagnostic plots suggested for the large majority of individual \heii emitters and the stacks of \emph{Faint} and \emph{Narrow} emitters. To compare with galaxies that do not show \heii emission, we randomly sample 43 spectra, the same number of sources that go into the stack of all \heii emitters, from the 849 galaxies with no \heii emission and stack their spectra. Choosing a comparable sample size results in comparable errors on the two stacks, which is essential for consistent metallicity determination. We also ensure that the spectra that are sampled have comparable stellar masses, SFRs, and redshifts to those with \heii emission. 

\begin{figure*}
    \centering
    \includegraphics[scale=0.8]{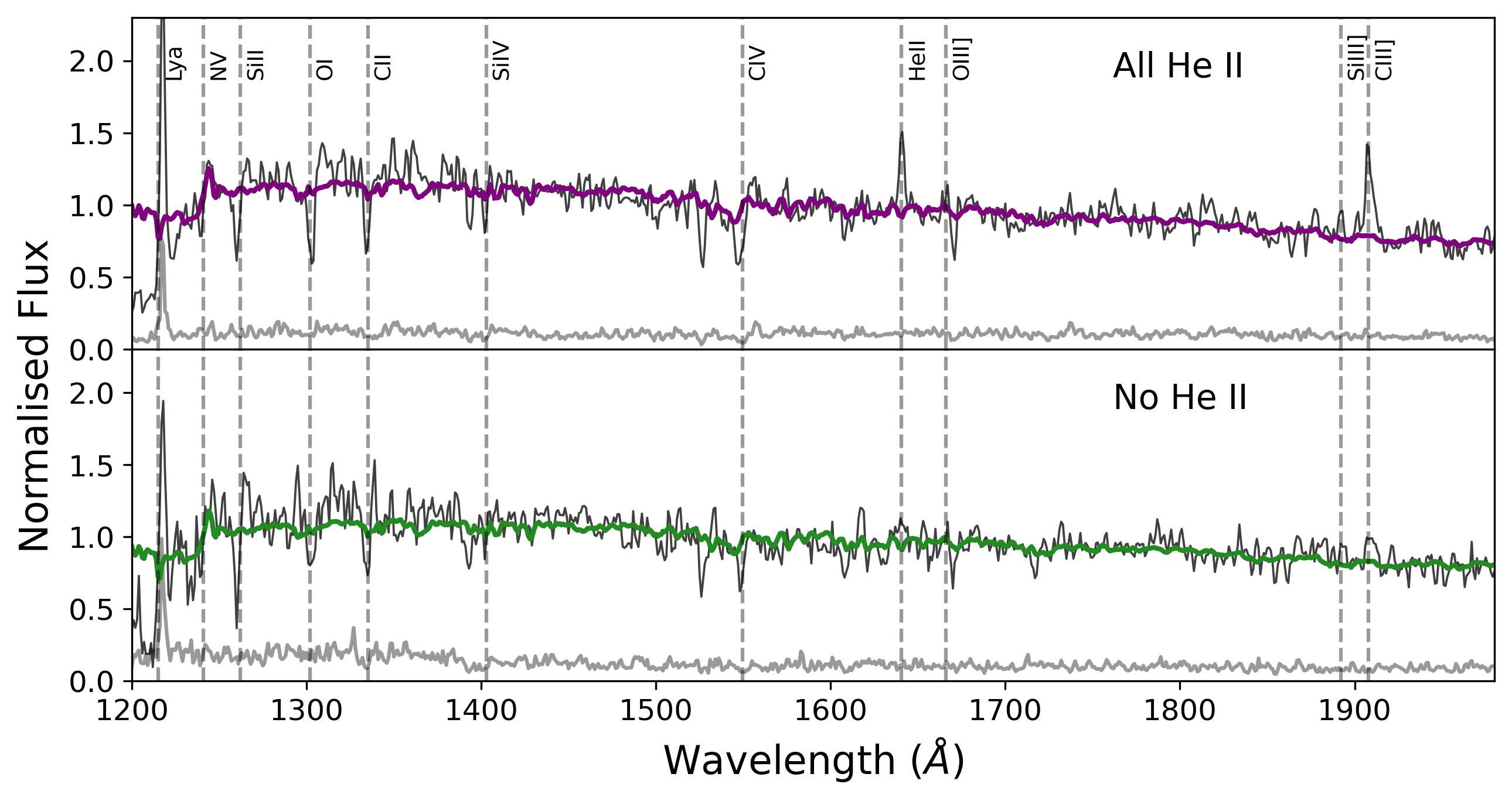}
    \caption{Best-fitting templates using the method of \citet{cul19} to determine the stellar metallicities from stacked spectra (shown in black, with errors in grey) of \heii emitters (\textit{top}) and non-emitters (\textit{bottom}). The metallicities derived for both classes of objects are highly comparable, at roughly 10\% that of solar metallicity. The inferred UV slope and fitted absorption features are comparable in both stacks, however the stack of \heii emitters has stronger emission lines compared to the stack of non-\heii emitters.}
    \label{fig:metallicity}
\end{figure*}
The metallicity that we measure for the stack of galaxies with no \heii is $Z=0.0020 \pm 0.0002$, and is consistent within the error bars with the stellar metallicity of \heii emitters. The best-fitting templates that are used to determine metallicities for the stacks of both \heii emitters and non-emitters are shown in Fig \ref{fig:metallicity}.  Similar to our result, \citet{ste16} and \citet{top19} also found that models with stellar metallicities of $0.1 Z_\odot$ provide the best match for their sample of star-forming galaxies, although their sample lies at slightly lower redshifts ($z\sim 2.4$) compared to our sample. Interestingly, our metallicity measurements show that although \heii -emitting galaxies have low metallicities, they do not have significantly lower metallicities than the general star-forming galaxy population at the same redshifts. Comparable metallicities between \heii emitters and non-emitters are indeed consistent with the other findings of this paper, where their derived physical properties (such as stellar mass, SFRs, rest-frame UV magnitudes, etc.) are not significantly different either. Therefore, from these results it is not immediately clear what causes only a small fraction of galaxies to show \heii emission. The UV continuum slope and absorption features visible in the two stacked spectra are also comparable, with the only difference being the presence of stronger UV emission lines in the stack of \heii emitters. We measure the UV slope, $\beta$, from the stacked spectra, where $F_\lambda \propto \lambda^\beta$, and find $\beta=-0.32$ for the stack of \heii emitters and $\beta=-0.30$ for the stack of non-emitters. The UV slope is shaped by the metallicity, age, and dust properties of the underlying stellar populations, and the fact that \heii emitters and non-emitters have very similar $\beta$ values reinforces the idea that the prevalent physical conditions in these two classes of galaxies are consistent.

Nebular \heii emission is most likely caused by ionisation due to massive stars residing within young stellar populations in galaxies. It has been shown that a combination of low-metallicity stellar populations, which contribute to the rest-frame UV optical lines, and a more evolved stellar population dominating the continuum emission can be used to explain the line ratios and SEDs of high-redshift galaxies \citep[see e.g.][]{sob15}. One possible way to describe the presence of \heii in some galaxies is that this emission (and other strong UV lines seen in the stacked spectrum) may be the result of a recent, extreme star-formation event in the galaxy. As a result, the bright UV emission lines in the stacked spectrum of \heii emitters, originating from young and massive stars (some possibly affected by interactions in binary systems), are superimposed on a more regular stellar continuum originating from the rest of the stars in the galaxy, which dominate the bulk of the stellar mass. As mentioned earlier, a caveat of the method used to estimate stellar metallicities means that if \heii -emitting galaxies indeed have young ages, then the measured metallicities will be biased. A direct probe for evidence of recent star-formation could be the H$\alpha$ emission line, and upcoming facilities such as the \emph{JWST} and \emph{E-ELT} will be instrumental in pioneering near-infrared spectroscopic studies of such faint galaxies.

\subsection{Alternative sources of \heii ionising photons}
In the following sections we briefly explore additional physical phenomena that could give rise to \heii emission in star-forming galaxies and have not yet been necessarily included in the stellar population synthesis models.

\subsubsection{X-ray binaries}
The difference between the \heii EWs predicted by stellar populations that include binary stars and the EWs that we observe in our sample suggests that there may be some additional mechanisms that could account for the missing He$^+$ ionising photons. X-ray binaries (XRBs) have been suggested as one such mechanism for several decades now \citep[see][for example]{gar91}. \citet{sch19} showed that XRB populations in low-metallicity galaxies can account for the \heii line strength, and \citet{for19} recently showed that the contribution from XRBs indeed increases with decreasing metallicities at high redshifts, as was already shown for low-redshift galaxies \citep{dou15, bro16}. Therefore, it is important to explore the role of XRBs in low-metallicity, \heii -emitting galaxies.

One of the VANDELS fields, the Chandra Deep Field South (CDFS), has the deepest X-ray data available, with an effective exposure time of 7 Ms. In a follow-up paper (Saxena et al. in prep), we will explore the X-ray properties of individual \heii -emitting sources in the CDFS that were identified in this paper, in addition to a stacking analysis of both \heii emitters and non-emitters with comparable physical properties. In combination with the latest models \citep{sch19, sen19b}, we will test whether XRBs are an important contributor to the \heii emission observed in low-metallicity galaxies across redshifts.

\subsubsection{Stripped stars}
It has been shown that interacting binary stars, which can result in one of the stars being stripped of its envelope, can emit a significant amount of ionising radiation -- He$^+$ ionisation in particular -- and can account for the missing ionising photons in stellar population synthesis models \citep{got18, got19}. The effect of stripped stars lasts for a much longer time after the initial starburst, and can provide He$^+$ ionising photons even at older stellar ages. However, the contribution towards He$^+$ ionising photons may not be enough even after the inclusion of stripped stars, as the ratio of the emission rate of ionising photons for \heii and H~\textsc{i}, $Q_2/Q_0 \sim 10^{-3.7}$, is still relatively low, for example, showed that $Q_2/Q_0 \sim 10^{-2}$ is needed to account for the \heii/ H$\beta$ observations.

\subsubsection{Wolf-Rayet stars}
Broad \heii emission has often been attributed to WR stars, both from theoretical \citep[see][for example]{sch96} and observational \citep{bri08} points of view. WR stars are rare, massive stars that have lost their outer hydrogen layer and are in the process of fusing helium or other heavier elements in their core. A characteristic feature of WR stars is the presence of broad emission lines due to fast stellar winds. \citet{sch96} showed that WR stars can power the broad \heii $\lambda4686$ emission line observed in galaxies. \citet{shi12} showed that WR stars in galaxies with low metallicities and with low stellar ages of $4-5$ Myr can explain the broad \heii emission, but a significant fraction of galaxies with nebular \heii emission do not show signatures of the presence of WR stars.

In our sample of \heii emitters, there are six galaxies that show broad \heii $\lambda1640$ (FWHM > 1000 km s$^{-1}$), which could likely be powered by WR stars. However, to confirm that WR stars are indeed powering these sources, looking for the presence of other spectral features associated with WR stars is necessary, such as WR bumps around \heii and \civ. In the spectra of sources with broad \heii in our sample, we examined the regions around the \heii and \civ UV lines but we do not see a bump in the spectrum around these regions. The presence of these bumps is not clear in the stacked spectra either. This is not surprising given the relatively low S/N of the spectra and the low number of galaxies that go into the stack. A full analysis searching for other WR features in the spectra of \emph{Broad} \heii emitters is beyond the scope of this work, and deeper observations of the rest-frame optical spectrum of these sources may hold some clues.

\subsubsection{AGNs}
Hard UV ionising photons produced by AGN accretion can provide the energies required to ionise \heii. Emission lines originating from AGNs are often very strong and depending on the viewing angle, could also be very broad (FWHM > 1000 km s$^{-1}$). Therefore, sources that show strong and/or broad \heii emission in our sample could be powered by AGNs. The UV line ratios of stacked spectra suggest that stacks of the \emph{Broad} \heii emitters indeed lie closer to the AGN regime than the star-forming regime (see Figure \ref{fig:gutkin_feltre}).  

Apart from the above, we do not see any clear or obvious signatures of AGNs (for example, the N \textsc{v} line) in the spectra of our galaxies. Strong \civ emission in \heii emitters has often been attributed to AGNs \citep[see][for example]{nan19}, however, as we showed in Section \ref{sec:civ_emitters}, BPASS models \citep{xia18} can explain the line ratios of galaxies with both \heii and \civ through star-formation alone, and indeed \civ has been observed in extremely low-metallicity galaxies with \heii emission in the local Universe.

Therefore, it is likely that sources with broad \heii emission could be powered by AGNs, but with currently available data it is not easy to be certain about the nature of the underlying ionising source. In a follow-up paper we will explore in greater detail the X-ray properties of our \heii emitters, and use the X-ray luminosities to confirm or rule-out the presence of AGNs in our sample.

\section{Conclusions}
\label{sec:conclusions}
In this paper we present \heii $\lambda1640$ -emitting galaxies in the redshift range $z=2.2-4.8$ selected from the VANDELS spectroscopic survey in the Chandra Deep Field South (CDFS) and the UKIDSS Ultra Deep Survey (UDS) fields. The sources were selected using a combination of visual examination of both 1D and 2D spectra, as well as emission-line-fitting techniques. Starting from a Parent sample of 949 spectra in the redshift range mentioned above in both fields with secure spectroscopic redshifts, we identified a sample of 33 \emph{Bright} (S/N > 2.5) \heii emitters and 17 \emph{Faint} (S/N < 2.5) \heii emitters.

For the \emph{Bright} \heii emitters, we performed careful measurements of the \heii emission line flux, FWHM, and EW. Where detected, we also measured other UV lines in the spectra of these galaxies. For the \emph{Faint} \heii emitters, we relied on stacking to infer their properties. We then analysed the physical properties of \heii emitting sources, both for individual galaxy spectra and stacks, and used UV emission line ratio diagnostics to set constraints on the underlying ionising mechanisms.

The main conclusions of this study are as follows:
\begin{itemize}
        \item The measured line fluxes of \emph{Bright} \heii emitters in VANDELS range from $1.1$ to $31.0\times10^{-18}$ \flux. The FWHM (rest-frame) range from $240$ to $2120$ km s$^{-1}$ and EWs (rest-frame) range from $0.9$ to $21.4$ $\AA$. The stellar masses inferred from SED fitting using broad-band photometry are in the range $\log_{10} M_\star=8.6-10.8$ $M_\odot$, UV corrected star-formation rates are in the range $\log_{10}(\textrm{SFR})=0.5-2.0$ $M_\odot$ yr$^{-1}$ , and rest-frame UV magnitudes (at 1500 $\AA$ are in the range $M_{\textrm{UV}}=-21.9$ to $-19.2$. When comparing the physical properties with the parent sample using KS tests, we do not see significant differences between galaxies that show strong \heii emission and the galaxies with no \heii emission in the parent sample.
        
        \item We identify seven possible AGNs in our sample of \heii emitters, which have been identified either due to the presence of strong \civ emission in their spectra, or an X-ray match in the publicly available catalogues. Additionally, we separated the \emph{Bright} \heii emitters that are not possible AGNs into subsamples based on the FWHM of their \heii line, and classify 20 sources as \emph{Narrow} (FWHM < 1000 km s$^{-1}$) and 6 sources as \emph{Broad} (FWHM > 1000 km s$^{-1}$). We then produced stacked spectra of narrow and broad emitters, in addition to a stacked spectrum of faint \heii emitters. We then used \heii, \oiii and \ciii lines, both from individual spectra and stacks to compare the observations with stellar population synthesis models. The models we use are from \citet{gut16} based on single stars, and from \citet{xia18} that take into account the effect of binary stars.
        
        \item The predictions from single star models suggest that the line ratios of a majority of our \heii emitters, including the stacks of \emph{Narrow} and \emph{Faint} emitters, can be explained by star-formation alone; however, there are degeneracies between the model parameters. Single star models are unable to reproduce the line ratios seen in \emph{Broad} \heii emitters, and we show that ionisation from AGNs is preferred for these. However, models including the effect of binaries from \citet{xia18} reproduce the line ratios of all \heii emitters in a more consistent manner when compared to single star models. The metallicities of the models that best match the observations are in the range $Z = 0.0001 - 0.002$ with stellar ages of $10^7-10^8$ yr. Our measured line ratios overlap almost completely with observations in the literature of \heii emitters at comparable redshifts, and compared to UV line ratios of local metal-poor dwarf galaxies, the metallicities required to explain our sources are generally lower.
        
        \item Although the low-metallicity binary star models are able to reproduce the UV line ratios of our \heii emitters, they underpredict the \heii EWs. \ciii EWs can be reproduced by low-metallicity models with ages below $10^7$ years, but the EWs of \heii are underpredicted by even the lowest metallicity models with young ages, as has been previously reported. This suggests that the inclusion of additional mechanisms capable of producing \heii ionising photons is required in stellar population synthesis modelling. Models with solar metallicity underpredict EWs of all UV lines considered in this study by a few orders of magnitude, leading to the conclusion that \heii emitters presented in this study are likely to have subsolar metallicities.
        
        \item We measure stellar metallicities on stacks of all \heii emitters (\emph{Bright} and \emph{Faint} combined) using features in the UV spectrum, and find a metallicity of $Z = 0.0022 \pm 0.0003$ $Z_\odot$, a value that is roughly 10\% of the solar metallicity. Interestingly, the metallicity measured for the stack of galaxies that do not show \heii emission is $Z=0.0020 \pm 0.0002$ $Z_\odot$, and consistent within error bars with that of \heii emitters. This suggests that galaxies that show \heii emission are not particularly metal poor when compared to the general galaxy population at these redshifts, but must have undergone a recent star-formation event that also boosts the strength of the other UV lines seen in their spectra. 
        
        \item We show that additional mechanisms such as a contribution from XRBs or stripped stars may be needed to properly reproduce the observed \heii EWs in the spectra of galaxies in our sample. In a follow-up paper, we aim to explore the X-ray properties of \heii emitters and set constraints on the XRB contribution in these galaxies.
\end{itemize}

\begin{acknowledgements}
      The authors thank the referee for useful and constructive comments that helped improve the quality of this manuscript. AS thanks Themiya Nanayakkara for useful discussions and for sharing the latest data that could be used for comparison in this paper. RS acknowledges support from the Amaldi Research Center funded by the MIUR program "Dipartimento di Eccellenza" (CUP:B81I18001170001). This work has made extensive use of \textsc{ipython} \citep{ipython}, \textsc{astropy} \citep{astropy}, \textsc{matplotlib} \citep{plt}, \textsc{mpdaf} \citep{mpdaf} and \textsc{topcat} \citep{topcat}. This work would not have been possible without the countless hours put in by members of the open-source developing community all around the world.  
\end{acknowledgements}


\bibliographystyle{aa}
\bibliography{HeII}

\appendix
\section{Rest-frame spectra of \emph{Bright} \heii emitters}
\begin{figure*}
        \centering
        \includegraphics[scale=0.8]{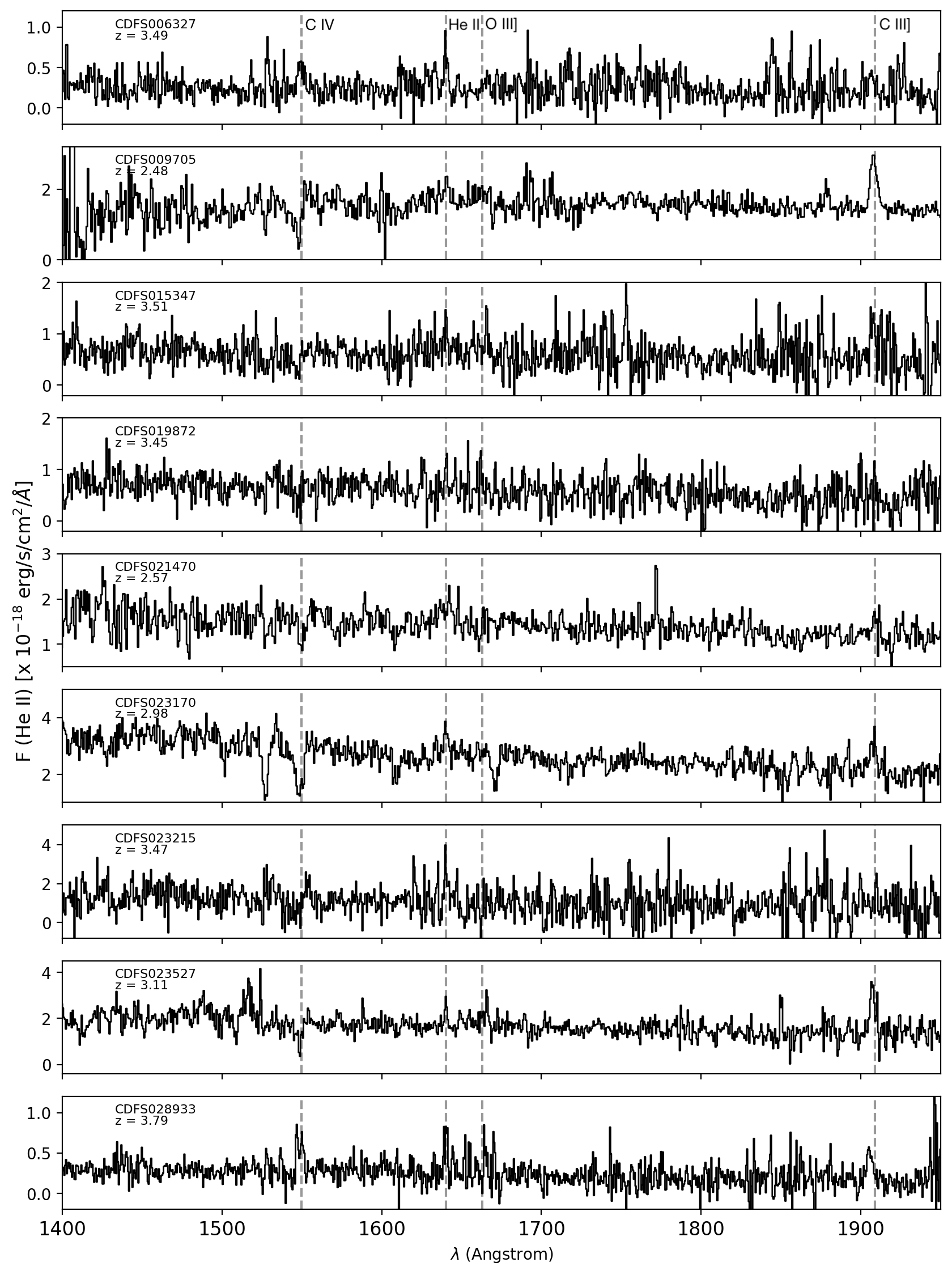}
        \caption{Rest frame spectra of \emph{Bright} ($\textrm{S/N}>2.5$) \heii emitters identified in the CDFS and UDS fields. Also marked in the spectra are other rest-frame UV emission lines used in this study.}
        \label{fig:CDFS_spectra}
\end{figure*}
\begin{figure*}
\ContinuedFloat
        \centering
        \includegraphics[scale=0.8]{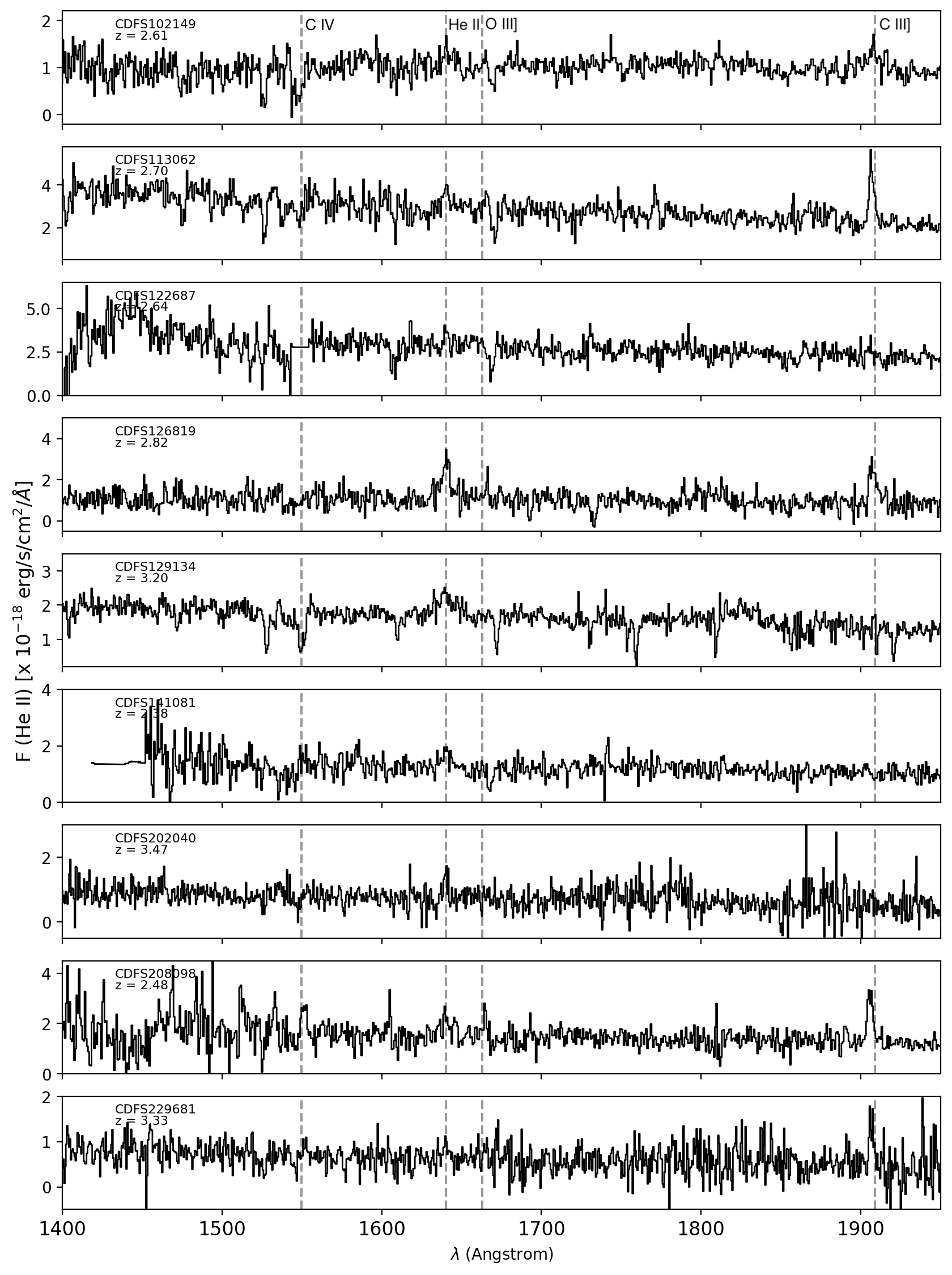}
        \caption{(continued)}
\end{figure*}
\begin{figure*}
\ContinuedFloat
        \centering
        \includegraphics[scale=0.8]{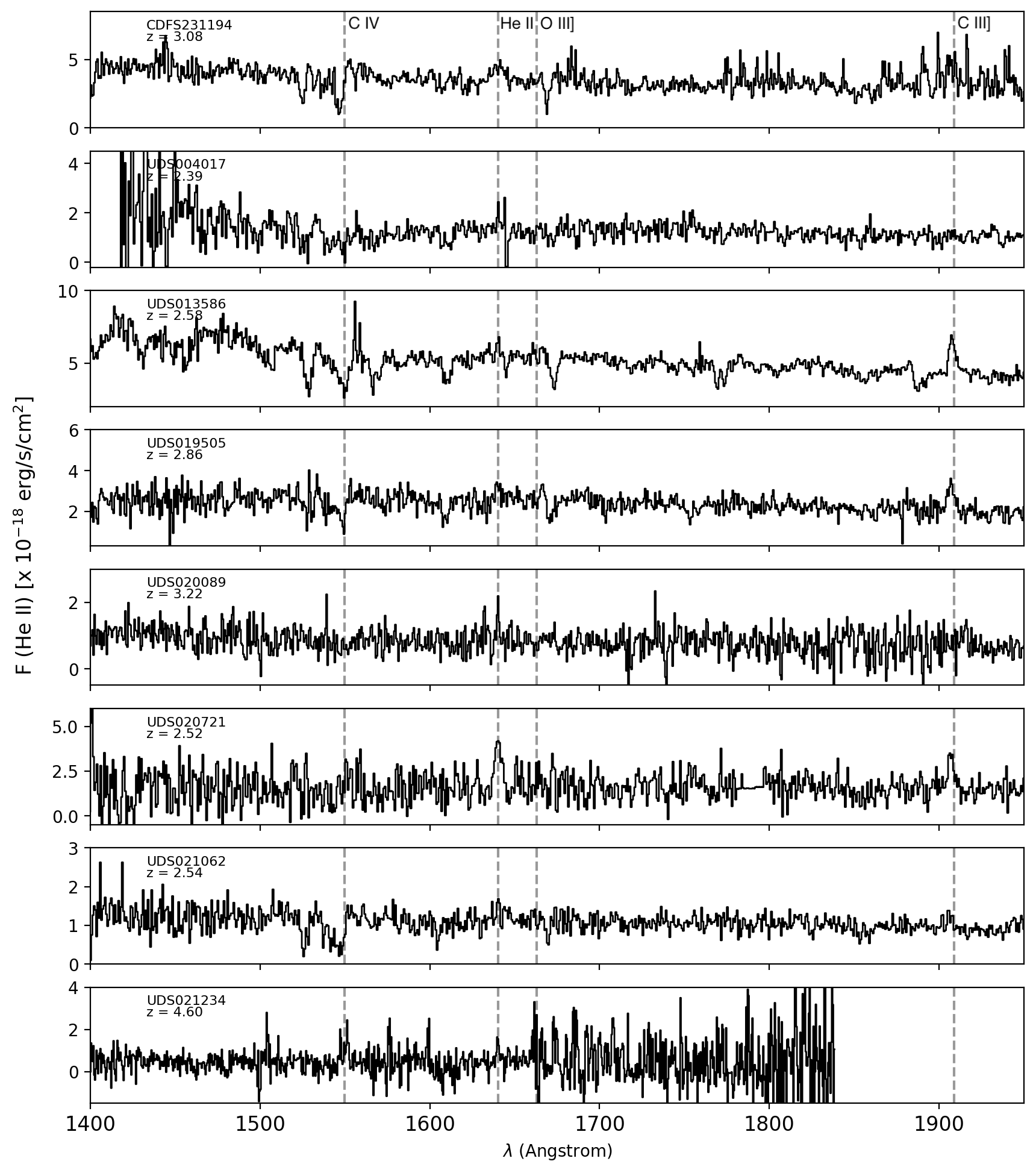}
        \caption{(continued)}
\end{figure*}
\begin{figure*}
\ContinuedFloat
        \centering
        \includegraphics[scale=0.8]{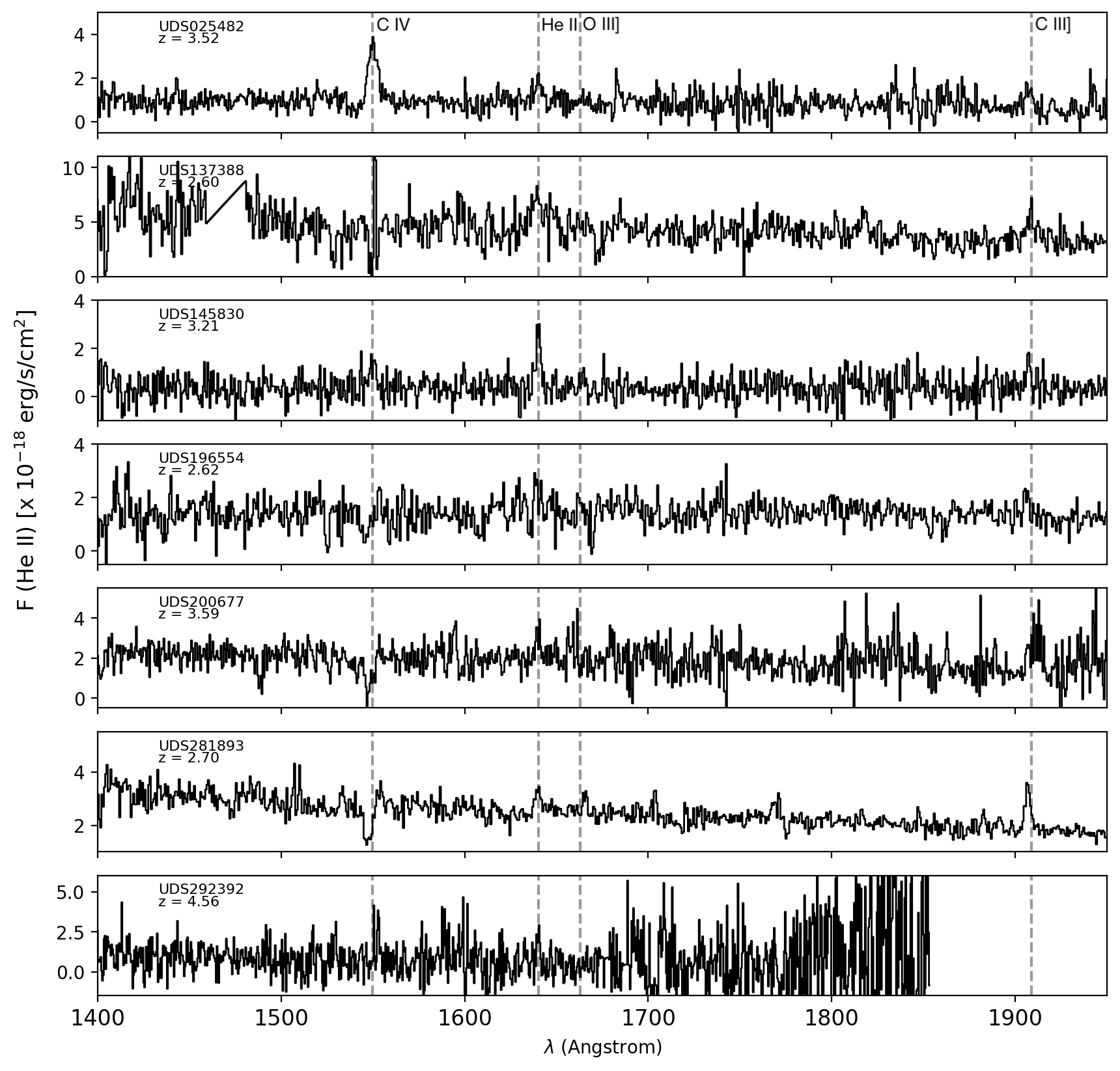}
        \caption{(continued)}
\end{figure*}

\end{document}